\documentclass[fleqn,usenatbib]{mnras}
\usepackage{newtxtext,newtxmath}
\usepackage[T1]{fontenc}
\DeclareRobustCommand{\VAN}[3]{#2}
\let\VANthebibliography\thebibliography
\def\thebibliography{\DeclareRobustCommand{\VAN}[3]{##3}\VANthebibliography}
\usepackage{graphicx}	
\usepackage{amsmath}	
\usepackage{arydshln}
\usepackage{caption}

\graphicspath{{./}{figures/}}
\title[VAST Orphan Afterglow Search]{A matched-filter approach to radio variability and transients: searching for orphan afterglows in the VAST Pilot Survey}
\author[J. K. Leung et al.]{James
K.\ Leung$^{1,2,3}$\thanks{E-mail: james.leung@sydney.edu.au}
Tara Murphy,$^{1,3}$
Emil Lenc,$^{2}$
Philip G. Edwards,$^{2}$
Giancarlo Ghirlanda,$^{4,5}$
\newauthor
David L.\ Kaplan,$^{6}$
Andrew O'Brien,$^{6}$
and Ziteng Wang$^{1,2,3}$
\\
$^{1}$Sydney Institute for Astronomy, School of Physics, The University of Sydney, NSW 2006, Australia\\
$^{2}$CSIRO Space and Astronomy, PO Box 76, Epping, NSW 1710, Australia\\
$^{3}$ARC Centre of Excellence for Gravitational Wave Discovery (OzGrav), Hawthorn, VIC 3122, Australia\\
$^{4}$INAF -- Osservatorio Astronomico di Brera, via E. Bianchi 46, I--23807 Merate (LC), Italy\\
$^{5}$INFN -- Sezione di Milano–Bicocca, Piazza della Scienza 3, 20126 Milano (MI), Italy\\
$^{6}$Department of Physics, University of Wisconsin-Milwaukee, P.O. Box 413, Milwaukee, WI 53201, USA\\
}

\date{Accepted 2023 May 27. Received 2023 May 22; in original form 2023 April 05}
\pubyear{2023}

\begin{document}
\label{firstpage}
\pagerange{\pageref{firstpage}--\pageref{lastpage}}
\maketitle

\begin{abstract}
Radio transient searches using traditional variability metrics struggle to recover sources whose evolution timescale is significantly longer than the survey cadence.
Motivated by the recent observations of slowly evolving radio afterglows at gigahertz frequency, we present the results of a search for radio variables and transients using an alternative matched-filter approach.
We designed our matched-filter to recover sources with radio light curves that have a high-significance fit to power-law and smoothly broken power-law functions;
light curves following these functions are characteristic of synchrotron transients, including ``orphan'' gamma-ray burst afterglows, which were the primary targets of our search.
Applying this matched-filter approach to data from Variables and Slow Transients Pilot Survey conducted using the Australian SKA Pathfinder, we produced five candidates in our search.
Subsequent Australia Telescope Compact Array observations and analysis revealed that: one is likely a synchrotron transient; one is likely a flaring active galactic nucleus, exhibiting a flat-to-steep spectral transition over 4\,months; one is associated with a starburst galaxy, with the radio emission originating from either star formation or an underlying slowly evolving transient; and the remaining two are likely extrinsic variables caused by interstellar scintillation.
The synchrotron transient, VAST J175036.1$-$181454, has a multi-frequency light curve, peak spectral luminosity, and volumetric rate that is consistent with both an off-axis afterglow and an off-axis tidal disruption event; interpreted as an off-axis afterglow would imply an average inverse beaming factor $\langle f^{-1}_{\text{b}} \rangle = 860^{+1980}_{-710}$, or equivalently, an average jet opening angle of $\langle \theta_{\textrm{j}} \rangle = 3^{+4}_{-1}\,$deg.
\end{abstract}

\begin{keywords}
galaxies: active -- radio continuum: transients -- gamma-ray bursts
\end{keywords}
\section{Introduction} \label{sec:intro}
In the standard fireball model, a gamma-ray burst (GRB) produces a panchromatic afterglow when the ultra-relativistic jet decelerates into the circumburst medium \citep{Meszaros1997}.
Owing to the relativistic beaming effect, the solid angle of the observable emitting region $\Omega$ increases over time as the bulk Lorentz factor $\Gamma$ of the relativistic blast wave decreases: $\Omega(t) \propto 1/\Gamma(t)^{2}$.

Orphan afterglows refer to those unaccompanied by an early, high-energy, prompt counterpart, which is beamed within the initial jet opening angle $\theta_{\textrm{j}}$.
They could arise from two scenarios: a ``dirty fireball'' origin or an off-axis viewing angle \citep{Rhoads2003}.
In the former scenario, a low initial bulk Lorentz factor $\Gamma_0$ (this is the Lorentz factor $\Gamma$ before the jet decelerates, corresponding to the prompt emission phase of the GRB) may prevent the escape of photons during the prompt emission phase as a consequence of large pair production opacity; however, it would still produce afterglow emission observable at longer wavelengths \citep[e.g.,][]{Dermer1999}.
In the latter scenario, the observer is viewing a classical afterglow off the jet axis, where the viewing angle $\theta_{\text{obs}}$ is beyond the jet opening angle $\theta_{\text j}$, i.e., $\theta_{\text{obs}} \geq \theta_{\text{j}} \gtrsim 1/\Gamma_0$ \citep{Rhoads1997}.
Here, the off-axis observer would not see the GRB prompt emission, but as the jet expands and decelerates to $\Gamma \approx 1/\theta_{\text{obs}}$, the panchromatic afterglow becomes visible from wider viewing angles. 

Distinguishing these two scenarios remains an observational challenge, given the similarities in their expected light curves and spectral properties \citep[e.g.,][]{Huang2002, Granot2018a}.
A well-sampled multi-wavelength light curve rise could provide the opportunity to distinguish these scenarios -- the presence of a fast X-ray transient, a shorter optical flux peak time, and early-time radio scintillation, would all provide evidence favouring the dirty fireball over the off-axis scenario \citep[e.g.,][]{Huang2002}.
These early-time light curves, however, are very difficult to obtain to the required sensitivity, especially when an orphan afterglow is discovered at longer wavelengths (since the emission may have already fallen below detectability thresholds at shorter wavelengths after its discovery).

Still, attempting to find orphan afterglows and subsequently distinguishing them between these two scenarios would provide useful insights into the properties of GRBs.
Detecting a sample of dirty-fireball afterglows could reveal whether they lie on a continuum with classical GRBs with high-baryon purity or whether there is a parametric dichotomy between the two classes, enhancing our understanding of the underlying physics of the progenitors \citep[e.g.,][]{Eichler2011}.
Studying off-axis afterglows could tighten constraints on the true rate of GRBs and the typical inverse beaming fraction, i.e., the ratio of all bursts to only those visible along the line-of-sight towards Earth, given as $\langle f^{-1}_{\text{b}} \rangle \cong 2 / \theta_{\text{j}}^2$ \citep[e.g.,][]{Frail2001}.
The population jet geometry could then be investigated by comparing empirical rates against predicted rates assuming different jet structures, e.g., top-hat jet \citep{Ghirlanda2014}, a universal structured jet \citep{Rossi2008}, or others.

Detections of any orphan afterglows in the past have been scarce owing to their faint flux levels.
The confirmation of any candidates has also been challenging due to the difficulty in distinguishing them from other slow transients; for example, supernovae and active galactic nuclei (AGNs) can often be sources of transient confusion in radio survey searches.
Despite these challenges, unconfirmed orphan afterglow candidates in radio survey searches \citep{Levinson2002, Gal-Yam2006} and non-detections of GRB radio counterparts in late-time follow-up of type Ibc supernova systems \citep{Soderberg2006} proved to be useful early on as they together allowed the typical inverse beaming fraction to be constrained to $60 < \langle f^{-1}_{\text{b}} \rangle < 10^4$.

Recently, improvements in search methods, instrumentation, and modelling have led to the discovery of likely orphan afterglow candidates. 
Photometric and spectral observations of SN 2020bvc pointed to the presence of a jet-cocoon\footnote{Signatures of a jet cocoon have previously been found in long GRB / supernova events \citep[e.g., GRB 171205A/SN 2017iuk;][]{Izzo2019}. In this scenario, the GRB jet is launched into the progenitor's stellar layers. The energy injection provided by the jet will give rise to a hot cocoon, which expands laterally to the jet itself. When the relativistic jet successfully penetrates the circumstellar material, it will produce a GRB and the standard afterglow emission. Meanwhile, the cocoon will continue to expand at mildly-relativistic velocities as it breaks out of the progenitor photosphere and will contribute additional thermal and non-thermal emission components.}, while the X-ray observations were consistent with an afterglow component; the emission was therefore attributed to a GRB viewed off-axis by ${\sim} 23\degr$, making this the first putative orphan afterglow discovery through an association with type Ic broad-line supernovae \citep{Ho2020a,Izzo2020}.
New unbiased optical surveys have yielded orphan afterglow candidates -- e.g., the Palomar Transient Factory discovery of PTF11agg \citep{Cenko2013}, a likely dirty fireball, and the Zwicky Transient Facility discoveries of 
ZTF20aajnksq/AT2020blt, ZTF21aaeyldq/AT2021any, and ZTF21aayokph/AT2021lfa (\citealt{Ho2022}; see also \citealt{Sarin2022}, \citealt{Gupta2022}, \citealt{Xu2023}, and \citealt{Lipunov2022}),
likely afterglows of on-axis GRBs missed by high-energy satellites. 
Modelling of X-ray transient CDF-S XT1 from the unbiased \textit{Chandra} Deep-Field South Survey showed the transient could be possibly interpreted as a slightly off-axis (${\sim} 10\degr$) short GRB orphan afterglow \citep{Sarin2021}.
While these high-energy and optical transient surveys are more effective for finding on-axis GRBs and dirty fireballs, they are less sensitive than radio transient surveys to the off-axis orphan afterglows beamed away from the observer at larger angles \citep[e.g.,][]{Frail2001, Chandra2012} and to events in dark dust-obscured regions \citep[e.g.,][]{Djorgovski2001}.

Previous wide-field radio surveys \citep[e.g.,][and references therein]{Mooley2016} lacked the sensitivity, sky coverage, and sampling cadence required to detect orphan afterglows and other extragalactic transients.
To overcome these challenges, the current generation of telescopes and design of unbiased transient surveys \citep[e.g.,][]{Murphy2013, Shimwell2017, Fender2016, Lacy2020} have incorporated various improvements that boosted their sensitivity towards extragalactic transients.
A comparison of sources in the first epoch of the VLA Sky Survey \citep[VLASS;][]{Lacy2020} with sources in the Faint Images of the Radio Sky at Twenty-cm \citep[FIRST;][]{Becker1995} survey has already led to the discovery of a decade-long extragalactic transient, FIRST J141918.9$+$394036 \citep{Law2018}.
Follow-up observations, including Very Long Baseline Interferometry (VLBI) and optical spectroscopy of the host galaxy, support the interpretation that the transient is likely an off-axis afterglow at a low redshift $z = 0.01957$, although an alternative interpretation of the transient as a nebula of a newly born magnetar has not been ruled out \citep{Marcote2019, Mooley2022}.

The Australian SKA Pathfinder \citep[ASKAP;][]{Johnston2007, Hotan2021} is an array of thirty-six 12\,m antennas located at Inyarrimanha Ilgari Bundara, the CSIRO Murchison Radio-astronomy Observatory, operating between 700 and 1\,800\,MHz. 
The ASKAP survey for Variables and Slow Transients \citep[VAST;][]{Murphy2013} is being conducted using this telescope, taking advantage of its large ${\sim} 30\,$deg$^{2}$ nominal field-of-view and capability of reaching 1\,mJy\,beam$^{-1}$ rms in 1\,min of integration.
Its ability to detect sources at the ${\sim} 1\,$mJy level over more than $10\,000\,$deg$^{2}$ makes it sensitive to discovering orphan afterglows \citep{Ghirlanda2014}, along with other extragalactic synchrotron transients, e.g., tidal disruption events (TDEs), type Ibc supernovae, etc.\ \citep{Metzger2015}.

One difficulty in detecting orphan afterglows and other extragalactic transients in gigahertz-frequency surveys is the slow evolution of their light curves \citep[see for example, the temporal decay of GRB 171205A as observed by ASKAP and uGMRT --][]{Leung2021,Maity2021}.
This is because emission from relativistic cosmic explosions peak at the gigahertz-band only at late-time when the blast wave has decelerated to mildly- or sub-relativistic speeds \citep[e.g.,][]{Chandra2012}.
Since traditional variability metrics, such as the reduced chi-square and modulation index \citep[e.g.,][]{Kesteven1976,Mooley2013,Rowlinson2019,Murphy2021}, are designed for finding variables or transients varying significantly on the timescales probed by the search, finding these slow transients in gigahertz variability surveys with observing cadence of a few months or shorter (which are the majority of such surveys) via a standard variability search will be challenging; this is discussed in more detail in \textsection\ref{sec:d+c}.
We have instead applied a matched-filter approach to finding slow transients in unbiased gigahertz surveys (also see \citealt{Feng2017}), which involves recovering sources with light curves having high-significance fits to smoothly broken power-law (SBPL) and power-law (PL) functions, characteristic of synchrotron transients, rather than just those with high epoch-to-epoch variability.

In this paper, we apply this method to search for orphan afterglows and other extragalactic synchrotron transients in the VAST Pilot Survey \citep[VAST-P;][]{Murphy2021}, spanning a total of 28 months from 2019 April 25 to 2021 August 24.
These observations and the pipeline we used for light-curve extraction are outlined in \textsection\ref{sec:observations}.
We detail our matched-filter search methodology for finding orphan afterglows in \textsection\ref{sec:search} and the follow-up observations/interpretation of the resulting candidates in \textsection\ref{sec:results}.
Finally, we conclude with discussing the possible implications of our results on both transient and GRB rates as well as for future radio transient studies in \textsection\ref{sec:d+c}.

In this paper, we assume a flat $\Lambda$-CDM cosmology with $H_0=~67.8$\,km\,s$^{-1}$\,Mpc$^{-1}$, $\Omega_{\text{M}} = 0.308$, and $\Omega_{\Lambda} = 0.692$ \citep{Planck2016}.

\section{Observations and Light-curve Extraction Pipeline} 
\label{sec:observations}
We used data from the first (VAST-P1) and second (VAST-P2) phases of the VAST Pilot Survey in our search for orphan afterglows and the subsequent follow-up of candidates. 
Using the VAST pipeline \citep{Pintaldi2022}, we associated sources across epochs in the data to build the light curves, which were the primary inputs of our search detailed in \textsection\ref{sec:search}. 

\subsection{VAST Pilot Survey Observations} \label{ssec:vastp}

\begin{table}
\caption{Table of VAST-P observations. 
Columns 1 through 5 show the epoch label, the number of fields in each epoch, the start and end dates for each epoch, and the total sky area covered for each epoch. 
Observations from VAST-P1 are given above the divider and observations from VAST-P2 are below the divider. 
Epochs 0 and 14 were conducted as part of RACS (only the subset of RACS observations overlapping with the VAST-P low- and mid-band footprints, respectively, are included), while epochs 15x and 16x were quality-gate observations conducted prior to the planned set of VAST-P2 observations. 
Epochs having labels suffixed with an `x' only cover a subset of the full footprint and epoch having labels suffixed with an asterisk (*) were taken at the mid-band ($1\,367\,$MHz image central frequency) while the others were taken at the low-band ($888\,$MHz image central frequency).
Our slow-transients pipeline run and orphan afterglow search only used the low-band observations, with the mid-band observations being used only in our candidate follow-up process.}
\label{tab:vast_epochs}
\begin{tabular}{ccccc}
\hline\hline
Epoch &  
\begin{tabular}[c]{@{}c@{}}No. of\\Fields\end{tabular} & 
\begin{tabular}[c]{@{}c@{}}Start Date\\(UT)\end{tabular} & 
\begin{tabular}[c]{@{}c@{}}End Date\\(UT)\end{tabular} & 
\begin{tabular}[c]{@{}c@{}}Sky Area\\(deg$^2$)\end{tabular} \\
\hline
0\phantom{00}      & 113 & 2019 Apr 25 & 2020 May 3 & \phantom{0}5\,006$^\dagger$ \\ 
1\phantom{00}      & 113 & 2019 Aug 27 & 2019 Aug 28 & 5\,131 \\
2\phantom{00}      & 108 & 2019 Oct 28 & 2019 Oct 31 & 4\,905 \\
3x\phantom{0}      &  43 & 2019 Oct 29 & 2019 Oct 29 & 2\,168 \\
4x\phantom{0}      &  34 & 2019 Dec 19 & 2019 Dec 19 & 1\,672 \\
5x\phantom{0}      &  81 & 2020 Jan 10 & 2020 Jan 11 & 3\,818 \\
6x\phantom{0}      &  49 & 2020 Jan 11 & 2020 Jan 12 & 2\,400 \\
7x\phantom{0}      &  33 & 2020 Jan 16 & 2020 Jan 16 & 1\,666 \\
8\phantom{00}      & 112 & 2020 Jan 11 & 2020 Feb 1 & 5\,097 \\
9\phantom{00}      & 112 & 2020 Jan 12 & 2020 Feb 2 & 5\,097 \\
10x                &  13 & 2020 Jan 17 & 2020 Feb 1 & 803 \\
11x                &  11 & 2020 Jan 18 & 2020 Feb 2 & 695 \\
12\phantom{0}      & 112 & 2020 Jun 19 & 2020 Jun 21 & 5\,100 \\
13\phantom{0}      & 104 & 2020 Aug 28 & 2020 Aug 30 & 4\,884 \\ 
\hline
$14^*$             &  90 & 2020 Dec 24 & 2021 Aug 1 & 2\,554 \\
15x                &   1 & 2021 Apr 1 & 2021 Apr 1 & 66 \\
\phantom{0}16x$^*$ &   2 & 2021 Apr 2 & 2021 Apr 2 & 67 \\
17\phantom{0}      & 113 & 2021 Jul 21 & 2021 Jul 24 & 5\,515 \\
$18^*$             &  91 & 2021 Jul 28 & 2021 Aug 1 & 2\,797 \\
19\phantom{0}      & 113 & 2021 Aug 20 & 2021 Aug 24 & 5\,122 \\
$20^*$             &  91 & 2021 Sep 20 & 2021 Sep 25 & 2\,604 \\
$21^*$             &  91 & 2021 Nov 18 & 2021 Nov 22 & 2\,604 \\
\hline
\end{tabular}

\raggedright
$^\dagger$ n.b. the sky area for the reference epoch 0 differs from that presented in \citet{Murphy2021} as our work here uses the subsequently publicly released images that were masked with a higher noise threshold
\end{table}

Table \ref{tab:vast_epochs} gives a detailed summary of the VAST-P observations across both Phase 1 and 2.
VAST-P1 consisted of 13 epochs (six full and seven partial\footnote{Epochs with partial coverage contain fields from extra test observations and from duplicate observations arising from problems in the scheduling of the full epochs.}) in addition to a reference epoch (0) conducted as part of the low-band Rapid ASKAP Continuum Survey \citep[RACS-low;][]{McConnell2020}.
Each field was observed at the $888\,$MHz low-band central frequency with an integration time of $12\,$min (with the exception of epoch 0 with $15\,$min integration time) and total bandwidth of $288\,$MHz, giving a typical rms sensitivity of $0.24\,$mJy\,beam$^{-1}$.
The full footprint of VAST-P1 covers ${\sim} 5\,100\,$deg$^{2}$ across 113 fields.
For details on the RACS data calibration and reduction pipeline, refer to \citet{McConnell2020}; \citet{Hale2021} for the RACS source catalogue and properties; and \citet{Murphy2021} for VAST-P data products and observing strategy.

VAST-P2 consisted of five full epochs (17-21), of which three were observed at the mid-band and the other two were at the low-band.
Preceding the full epochs were two quality-gate epochs (15x and 16x) conducted as a test prior to the planned set of VAST-P2 observations and a reference epoch (14) for the mid-band conducted as part of the mid-band Rapid ASKAP Continuum Survey \citep[RACS-mid;][]{Duchesne2023}.
The low-band observations followed the same observational parameters as in VAST-P1, while the mid-band observations used a smaller footprint and were centred on a central frequency of $1\,296\,$MHz with $12\,$min integration time per field (with the exception of epoch 14, which had a $15\,$min integration time).
However, since the lower $144\,$MHz of the band is flagged due to radio frequency interference, the resulting images have a central frequency of $1\,397\,$MHz and a bandwidth of $144\,$MHz.
The full footprint of VAST-P2 mid-band covers ${\sim}2\,600\,$deg$^{2}$ across 91 fields.
Please refer to the references above and references therein for details relating to the data processing and resulting data products for VAST-P2 as they use a similar data reduction pipeline to that used for VAST-P1.

Following the procedures in \citet{Murphy2021}, we evaluated the astrometric accuracy and flux-density scale of our dataset\footnote{The only difference in procedure was the definition of compactness used for the mid-band data, which was defined in this work as $0.8 < S_I / S_P < 1.2$, since the fitted `envelope' function from \citet{Hale2021} for characterising whether a source was unresolved is only applicable to low-band ASKAP data.}; we provide a summary in Table \ref{tab:qcsummary}. 
The VAST-P sources in the low-band data had an astrometric offset with respect to the positions of associated sources in the International Celestial Reference Frame \citep[ICRF;][]{Charlot2020} catalogue of $0\farcs58 \pm 0\farcs51$ in right ascension and $-0\farcs32 \pm 0\farcs55$ in declination, with a standard error of 0\farcs01 in both coordinates (the standard error on the \textit{median} offset is given by $1.253 \times \sigma_\textrm{rms}/\sqrt{N}$, where $\sigma_\textrm{rms}$ is the rms spread of the measured offsets and $N$ is the total source count; e.g., \citealt{Maindonald2006}). 
For the mid-band data, this was $0\farcs28 \pm 0\farcs88$ in right ascension and $-0\farcs33 \pm 0\farcs49$ in declination, with a standard error of $0\farcs05$ in both coordinates.
The overall median VAST-P / RACS-low flux-density ratio for the low-band data was $1.04 \pm 0.17$, with all 16 epochs having a median flux-density ratio within 5 per cent of the overall median.
For the mid-band data, this was $0.96 \pm 0.28$, with all 5 epochs having a median flux-density ratio within 1 per cent of the overall median.
While RACS-low is at a lower frequency than the mid-band data, it was still used as the reference catalogue as its flux-density scale is well characterised in \citet{McConnell2020} and covers the entirety of the VAST-P mid-band footprint.
We corrected for the frequency difference in our flux-density ratio calculations by scaling with an assumed spectral index $\alpha$ of $-0.8$ \citep{Condon1992}, defined as $S_\nu \propto \nu^{\alpha}$.
The median astrometric offsets, each smaller than the size of an image pixel, were sufficient for robust cross-epoch source association for light-curve extraction, while the flux-density scale, consistent with all other epochs to within 5 per cent, was sufficient for the use of functional fits for our search method.

\begin{table}
	\centering
	\caption{Summary of the astrometric accuracy and flux-density scale of the VAST-P dataset used in this work. Columns 1 through 4 show the survey observing band, median right ascension and declination offsets for VAST-P sources with respect to ICRF sources, and the flux-density ratio of VAST-P sources compared against sources in the reference RACS-low catalogue. 
	We assumed a spectral scaling of $\alpha = -0.8$ to evaluate the flux-density ratio for mid-band data against the reference RACS-low catalogue.
	}
	\label{tab:qcsummary}
	\begin{tabular}{lccc} 
		\hline
		\hline
		VAST-P Band &
		\begin{tabular}[c]{@{}c@{}}RA Offset\\(arcsec)\end{tabular} & 
		\begin{tabular}[c]{@{}c@{}}Dec Offset\\(arcsec)\end{tabular} & 
		$S_{\text{VAST-P}}/S_{\text{RACS}}$\\
		\hline
		Low (888\,MHz) & $+0.58 \pm 0.51$ & $-0.32 \pm 0.55$ & $1.04 \pm 0.17$ \\
		Mid (1\,397\,MHz) & $+0.28 \pm 0.88$ & $-0.33 \pm 0.49$ & $0.96 \pm 0.28$ \\
		\hline
	\end{tabular}
\end{table}

For our light-curve extraction and subsequent search for orphan afterglows, we used the low-band data (17 of the 22 epochs) only to minimise uncertainties associated with spectral scaling.
We used the mid-band data only in the follow-up process to provide supplementary spectral information for characterising candidates in \textsection\ref{sec:results}.
This gives our search a temporal baseline of 28 months, spanning 2019 April 25 to 2021 August 24, with sampling cadences ranging from 1 day to 8 months.

\subsection{VAST Pipeline Data Analysis} \label{ssec:pipeline}

The VAST pipeline takes input images and source catalogues to produce light curves for every source that had a detection in one or more epochs.
Sources were associated between epochs using the de Ruiter radius, which is a quantity representing the angular separation between two sources normalised by their positional errors \citep{deRuiter1977,Scheers2011}, expressed as: 
\begin{equation}
    r_{ij} = \sqrt{\frac{(\Delta\alpha_{ij} \textrm{cos}\bar{\delta}_{ij})^2}{\sigma^2_{\alpha,i} + \sigma^2_{\alpha,j}} + \frac{(\Delta\delta_{ij})^2}{\sigma^2_{\delta,i} + \sigma^2_{\delta,j}}},
\end{equation}
where $\Delta \alpha (\delta)_{ij}$ is the positional offset in right ascension (declination) between sources $i$ and $j$, $\bar{\delta}_{ij}$ is the mean declination of sources $i$ and $j$, and $\sigma_{\alpha(\delta),i}$ is the $1\sigma$ positional uncertainty for source $i$ in right ascension (declination).
The resulting light curves after source association consisted of flux-density measurements from source catalogues produced using {\sc Selavy} \citep{Whiting2012} for detections and forced extractions\footnote{The flux density was measured at a specified location using the raw image, rms and background maps with the following package:\\ \url{https://github.com/askap-vast/forced_phot}} for non-detections.
The pipeline performed these forced extractions for each source in every epoch where there was a non-detection to build a complete light curve.
From these light curves, the pipeline calculated key variability metrics for each source -- the modulation index, $V$; and the reduced $\chi^2$ relative to a constant model, $\eta$ \cite[see also][]{Swinbank2015, Rowlinson2019}.
They measure the degree and significance of variability, respectively, and are defined as:
\begin{equation}\label{eq:var1}
V = \frac{1}{\overline{S}}\sqrt{\frac{N}{N-1} (\overline{S^2} - \overline{S}^2)},    
\end{equation}
\begin{equation}\label{eq:var2}
\eta = \frac{N}{N-1}\left( \overline{w S^2} - \frac{\overline{w S}^2}{\overline{w}}\right),
\end{equation}
where $N$ is the number of data points in the light curve, $\overline{S}$ is the mean flux density across all epochs, and $\overline{w}$ is the mean of the measurement weights across all epochs, defined as $w_i=1/\sigma_i^2$ with $\sigma_i$ the measurement uncertainty at the $i$-th epoch.
For more details on the VAST pipeline, please refer to \citet{Pintaldi2022}.

Our pipeline run ingested all the low-band epochs of VAST-P (i.e., epochs 1-13, 15x, 17, 19). 
The pipeline run settings followed those described in \citet{Murphy2021}, with the de Ruiter association radius being the key setting that controlled the source association and light-curve building process for this work.
We set the de Ruiter association radius parameter to 5.68, which effectively allows only $10^{-7}$ genuine associations to be missed\footnote{This is based on the properties of the Rayleigh distribution. The de Ruiter positional differences between genuinely associated sources will follow this distribution \citep[for details, see][]{deRuiter1977}.} and, in practice, this would correspond to an angular association radius that is much smaller than the size of our beam.
This pipeline run returned 1\,068\,985 unique sources.
We then applied the following filters to the dataset, including only sources that meet the following criteria:
\begin{enumerate}
    \item has at least four measurements (either forced or {\sc Selavy}) as required by our methods in \textsection\ref{sec:search};
    \item is compact as defined in \citet{Hale2021} by the integrated-to-peak flux-density ratio: 
    ${S_I / S_P < 1.025 + 0.69 \times \textrm{SNR}^{-0.62}}$;
    \item is isolated with no neighbouring sources within a $<1$\arcmin\ separation radius to avoid source and sidelobe confusion;
    \item has a minimum SNR of 10 for sources with one detection to minimise false detections (we relaxed this requirement for sources with more than one detection, requiring a lower minimum SNR of 7.5) -- see \citet{Metzger2015} for a brief discussion;
    \item has a median image rms $< 0.8\,$mJy\,beam$^{-1}$ to avoid regions of high noise, e.g., at the footprint edge;
    \item has a positive modulation index $V$, excluding negative flux-density artefacts caused by bright sources.
\end{enumerate}
After applying these filters, our sample included 130\,406 unique sources with at least one detection in a low-band VAST-P epoch.
A standard search for variables and transients could be performed at this point by identifying sources with large $V$ and $\eta$ values exceeding a certain threshold \citep[e.g., similar to][]{Murphy2021}.
However, as this method is prone to missing slow transients (discussed in more detail in \textsection\ref{sec:d+c}), we developed a more suitable method for our search for orphan afterglows.

\section{Orphan Afterglow Search} \label{sec:search}
In our search, we identified sources from our sample which had light curves featuring power-law rises and/or decays; these are characteristic of GRB afterglows, and more generally, extragalactic synchrotron transients \citep[e.g.,][]{Sari1998}. 
For each unique source in our sample, we defined its \textit{event duration} to range from the time of the first detection until the time of the first subsequent non-detection, where a detection was defined as a measurement with a SNR $\geq 2$.
In the case where the last measurement was also a detection, we instead counted until the time of the last detection.
We justify that we have improved our confidence in our low-signal detections (those with SNR from 2 to 7.5) by ensuring that these are spatially associated with at least two $>7.5\sigma$ detections and/or at least one $>10\sigma$ detection from other epochs (as discussed previously in \textsection\ref{ssec:pipeline}).
We conducted our search by performing a functional fit to all measurements within the event duration for every unique source in our sample, where the functional form depended on the number of observations $n$ taken within the event duration and the location of the highest flux-density measurement, i.e., the \textit{peak measurement}.

For sources with $n \geq 6$ and a peak measurement not located as the first or last measurement in the event duration, we fitted a SBPL model:
\begin{equation}
    \label{eq:sbpl}
    S_t = S_{\rm{peak}}
    \Bigg[\bigg(\frac{t-\Delta t}{t_{\rm peak}}\bigg)^{-s\delta_1} + \bigg(\frac{t-\Delta t}{t_{\rm peak}}\bigg)^{-s\delta_2}\Bigg]^{-1/s},
\end{equation}
where $t$ is the time after the first measurement (epoch 0), $S_t$ is the flux density as a function of time $t$, 
$\Delta t$ is the time elapsed between the time of the first measurement (epoch 0) and the time of the GRB explosion (or transient event), 
$\delta_1$ is the asymptotic rise slope, $\delta_2$ is the asymptotic decay slope, $S_{\rm{peak}}$ and $t_{\rm{peak}}$ are the approximate flux density and time post-burst of the light-curve peak, and $s$ is the smoothness parameter.
Since the light curve is often sampled at $> 1\,$month cadence, we chose to reduce the number of measurements required for a fit by reducing the number of free parameters in this model to increase the sensitivity of our search to transients evolving on the timescale of a few months.
We did this by fixing $s$ to the fiducial value of 5 and did not consider additional breaks in our functional form, which may more aptly describe GRBs with observed jet breaks \citep{Rhoads1999,Sari1999} and those exploding in a stellar-wind environment \citep{Chevalier2000}.

For the scenarios where (a) the source has $n \geq 4$ and a peak measurement located as either the first or last measurement in the event duration, or (b) the source has $n = 5$ and a peak measurement that is not the middle measurement, we fitted a simple PL model:
\begin{equation}
    \label{eq:pl}
    S_t = A (t-\Delta t)^\beta,
\end{equation}
where $\beta$ is the temporal slope (positive for rise and negative for decay), $A$ is a normalisation constant, and $t$, $S_t$, $\Delta t$ are defined as in Equation \ref{eq:sbpl} above.
In scenario (b) where not enough data points were available to fit a SBPL, we fitted the simple PL only to observations on the side of the peak measurement with the most data points.

We performed these functional fits using the non-linear least-squares optimisation routine in the {\sc SciPy} \citep{SciPy2020} {\sc Python} library. We applied the following physically motivated constraints during the fitting process:
\begin{enumerate}
    \item $ 0.8\,\text{max}(S_t) < S_{\text{peak}} < 3\,\text{max}(S_t)$ -- this constraint helps to exclude sources with erroneous fits of the SBPL peak, especially in cases where the light curve around the peak is undersampled, where as a result it becomes difficult to characterise the true properties of the underlying light curve;
    \item $ 10\,\text{d} < t_{\text{peak}} < 1\,000\,\text{d} $ -- afterglows typically peak at ${\sim} 100+$ days post-burst at gigahertz frequency and the constraints used here are consistent with known radio-afterglow light curves at this frequency \citep[e.g.,][]{Chandra2012, Maity2021};
    \item $-30\,\text{yr}\,< \Delta t < \text{min}(t)$ -- this constraint allows the search to be sensitive to bursts occurring up to $30$\,yr ago, accounting for the possibility of detecting decade-long transients, such as FIRST J141918.9$+$394036 \citep{Law2018};
    \item $0.2 < \delta_1 < 10$, $-5 < \delta_2 < -0.3$, and $-5 < \beta < 10$ -- numerous analytical and numerical efforts \citep[e.g.,][]{Granot2002, vanEerten2011, Granot2018b, Granot2018a, Lamb2021} have attempted to model radio light curves of afterglows seen off-axis.
    Their findings show the rise and decay behaviours differ substantially from the standard on-axis GRB afterglow \citep[e.g.,][]{Sari1998} due to complex interplay of jet dynamics with viewing angle effects; the typical rise and decay slopes varied based on a range of factors including but not limited to the viewing angle, microphysics (e.g., stratification parameter), and assumptions about the jet geometry (e.g., top-hat, Gaussian, etc.).
    Our constraints on these slopes allow our search to be sensitive to the most extreme temporal indices predicted.
\end{enumerate}
To measure how well the light curves were described by these functional fits, we calculated their corresponding $\chi^2$-statistic:
\begin{equation}
    \label{eq:chi2}
    \chi^2 = \sum^{n}_{i=1} \bigg(\frac{S_{i} - \widehat{S_{i}}}{\sigma_{i}} \bigg)^2,
\end{equation}
where $S_i$, $\sigma_i$ and $\widehat{S_i}$ are the measured flux density, the $1\sigma$ uncertainty on the measurement, and the model-fitted flux density at the $i$-th epoch of the $n$-epoch light curve, respectively.
This quantity is distributed according to the $\chi^2$-distribution with $n - p$ degrees of freedom, where $n$ is the number of measurements in the light curve and $p$ is the number of free parameters in the functional fit.

Aside from fitting each unique source in the sample to (SB)PL models, we also compared these fits against a constant benchmark model $S_t = \rm{constant}$, for which we also calculated the $\chi^2$ statistic.
Persistent sources that are not varying, as well as other variable and/or transient phenomena that exhibit oscillatory, burst- or pulse-like light curves, such as AGN scintillation or flaring radio stars, would be better described by this benchmark model than the (SB)PL functions.
Given the benchmark model is `nested' within the (SB)PL models, we determined which light curves were \textit{significantly} better described by (SB)PL models than the benchmark by calculating the $F$-statistic \citep[e.g.,][]{Weisberg2005}:
\begin{equation}
    \label{eq:F}
    F = \frac{\chi^2_{\rm b} - \chi^2_{\rm m}}{p_{\rm m} - p_{\rm b}}
    \bigg/ \frac{\chi^2_{\rm m}}{n - p_{\rm m}}
    = \bigg(\frac{\chi^2_{\rm b}}{\chi^2_{\rm m}} - 1 \bigg)
    \bigg/ \bigg(\frac{p_{\rm m} - p_{\rm b}}{n - p_{\rm m}}\bigg)
    ,
\end{equation}
where $\chi^2$ is calculated from Equation \ref{eq:chi2}, $p$ is the number of model free parameters, $n$ is the number of measurements, and the subscript $m$ refers to the (SB)PL models and the subscript $b$ refers to the benchmark model.
This quantity is distributed according to the $F$-distribution with ($p_{\rm m}$--$p_{\rm b}$,\,$n$--$p_{\rm m}$) degrees of freedom.

Having fit all the light curves and calculated the $\chi^2$-, $F$-statistics, we discarded sources from our sample which had:
\begin{enumerate}
    \item $p$-value$(\chi^2_{\text m})$ $< 0.05$ significance level -- the (SB)PL models do not fit the light curve \textit{significantly} well;
    \item $p$-value$(F)$ $> 0.05$ significance level -- the (SB)PL models do not fit the light curve \textit{significantly} better than the benchmark model;
    \item fitted parameters with values at the constraint boundaries, e.g., $\widehat{\delta_1} \approx 10$ -- the optimisation algorithm suggests the light curve would be better fit with unphysical parameter values;
    \item $|\beta| < 0.3$ -- the slow rise or decay is consistent with a flat, non-varying light curve;
    \item $\beta > 0$ and $\Delta t < -500$ days -- the light curve has not turned over despite bursting more than 500 days prior. 
    Turnovers occurring later than 500 days post-burst at gigahertz frequency have not been observed in large radio-afterglow samples \citep[e.g.,][but also see \citealt{Ghirlanda2014}]{Chandra2012}.
\end{enumerate}
After applying these cuts based on the functional fits, 193 (109 and 84 from the SBPL and PL fits, respectively) candidates remained in our sample. Beyond using our model-fitting methodology, we used additional radio and multi-wavelength data\footnote{The association radius for sources detected in archival radio surveys and by the \textit{Wide-field Infrared Survey Explorer} (\textit{WISE}) was 5\arcsec.
This association radius factored in the typical astrometric uncertainties in our ASKAP data (see Table \ref{tab:qcsummary}) in addition to the typical astrometric uncertainties of the archival/\textit{WISE} data (see their respective papers for details), which was often worse due to a significantly larger point spread function.
However, a smaller association radius of 3\arcsec (still conservative, $\gtrapprox 3\sigma_\textrm{ASKAP}$) was used for crossmatches with the optical surveys since the astrometric uncertainties were dominated by the ASKAP data in this case.}
as detailed below to further narrow down our candidate list. The candidate cuts we made include sources with:
\begin{enumerate}
    \item Detections in archival radio surveys, indicating that they are persistent sources or variable sources. The exception is for sources fitted with a negative decay slope $\beta < 0$ using a PL model, in which case, the time elapsed between the burst and the first detection $(- \Delta t)$ was also checked. The archival radio surveys we had cross-correlated with included the NRAO VLA Sky Survey \citep[NVSS;][]{Condon1998}, Sydney University Molonglo Sky Survey \citep[SUMSS;][]{Mauch2003}, Australia Telescope 20 GHz survey \citep[AT20G;][]{Murphy2010}, GaLactic and Extragalactic All-sky Murchison Widefield Array survey \citep[GLEAM;][]{Hurley-Walker2017}, and TIFR GMRT Sky Survey \citep[TGSS;][]{Intema2017}. 
    This archival data ruled out another 85 candidates (54 SBPL, 31 PL).
    \item Spectral or temporal information from supplementary radio data that was inconsistent with the expected behaviour of afterglows and synchrotron transients. We used supplementary data from VLASS, mid-band data from VAST-P2 and publicly available archival ASKAP data from the CSIRO ASKAP Science Data Archive \citep[CASDA;][see Data Availability statement for more detail]{Chapman2017,Huynh2020}. A typical example of this involves a light curve exhibiting a power-law decline at a lower frequency, yet is rising at a higher frequency at a later time. 
    We used supplementary data to rule out a further 43 candidates (26 SBPL, 17 PL).
    \item Spatially consistent (i.e., not offset from nucleus) \textit{Wide-field Infrared Survey Explorer} \citep[\textit{WISE};][]{Wright2010} counterparts, with infrared colours suggesting the source is likely a galaxy or an AGN. 
    In particular, we relied on source classifications based on the $[3.4\,\textrm{\textmu m}] - [4.6\,\textrm{\textmu m}]$ and $[4.6\,\textrm{\textmu m}] - [12\,\textrm{\textmu m}]$ infrared colours, following the colour-colour classification system presented in \citet{Wright2010} (see Figure~\ref{fig:wise_cc}). 
    Sources with infrared counterparts associated with AGNs were immediately ruled out since GRBs, with the exception of short GRB 150101B \citep{Xie2016} and long GRB 191019A \citep[interpreted to have been produced by the merger of a compact binary formed via dynamical interactions;][]{Levan2023}, are not known to be hosted by such systems. 
    We made these cuts by using the AllWISE catalogue and its associated data products \citep{Cutri2014}.
    This data allowed us to rule out a further 40 candidates (13 SBPL, 27 PL).

\begin{figure}
	\includegraphics[width=\linewidth,clip,trim={1mm 1mm 1mm 1mm}]{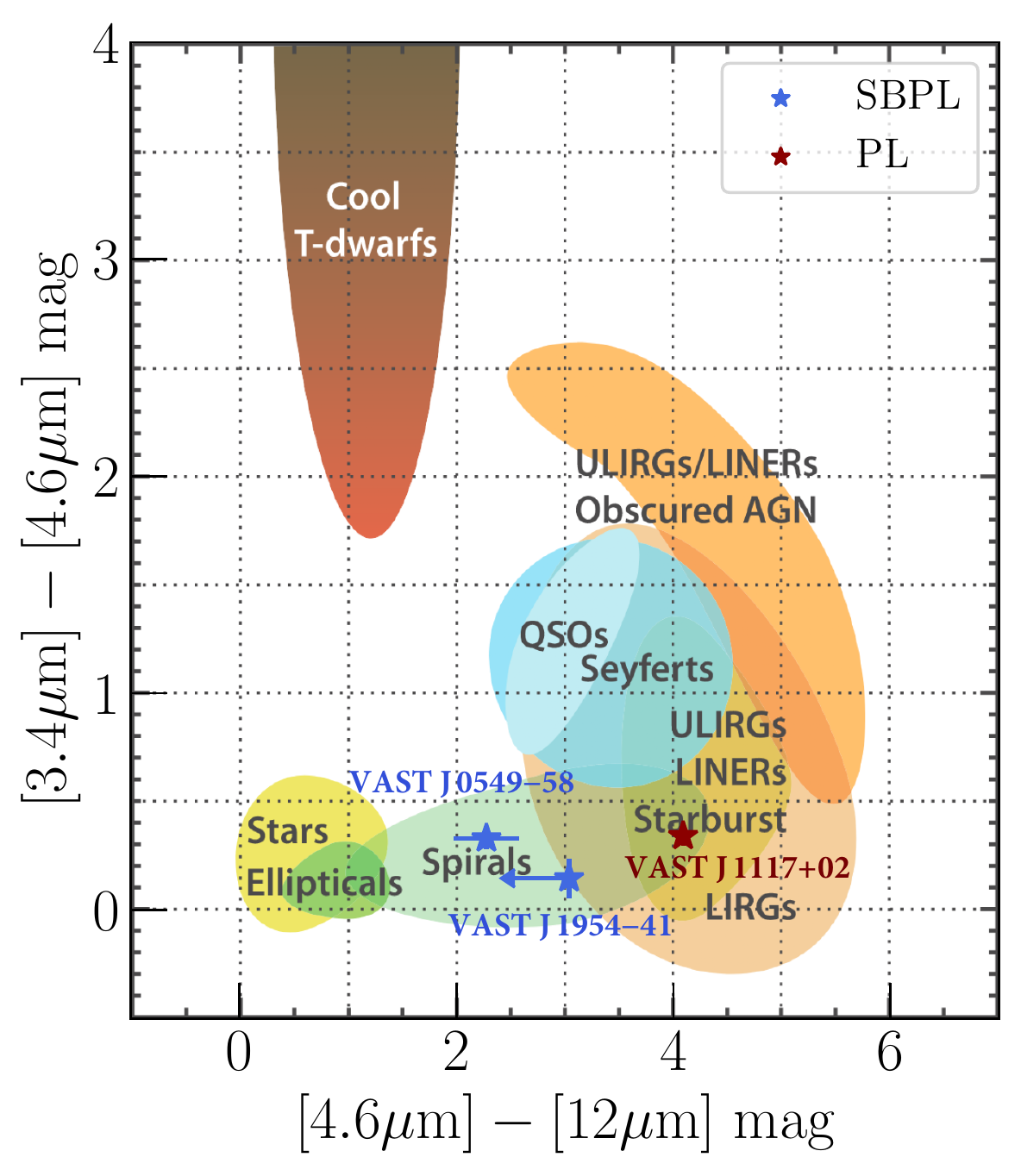}
    \caption{
    Orphan afterglow candidates obtained from our functional fits are overlaid on top of the \textit{WISE} colour-colour classification regions described in \citet{Wright2010}.
    Only candidates with a \textit{WISE} counterpart are shown.
    Candidates that were obtained in the search with a SBPL fit are shown in blue, while those obtained with a PL fit are shown in red.
    We note that for VAST J195443.9$-$412511 an upper limit is given for the $[4.6\,\textrm{\textmu m}] - [12\,\textrm{\textmu m}]$ colour due to a non-detection in the $12\,\textrm{\textmu m}$ band.
    }
    \label{fig:wise_cc}
\end{figure}
    
    \item Other reasons indicating they were likely not afterglow related. These reasons included: data artefacts (e.g., incorrect dates extracted for the light curve or bright nearby sources affecting image quality), photometric redshifts derived from the Dark Energy Camera Legacy Survey \citep[DECaLS;][]{Dey2019,Zhou2021} with $z \gg 0.2$ \citep[corresponding to the radio afterglow detectability threshold for VAST-P as shown in][]{Leung2021}, inconsistent peak and integrated flux-density light curves for borderline extended sources (slightly below the compactness threshold), deviations from the power-law fits greater than that expected from scintillation\footnote{The scintillation model we used here is detailed later in \textsection\ref{ssec:candidates}. We excluded sources with a modulation index (after accounting for the variability that can be explained by the power-law fits) exceeding that predicted by the scintillation model by more than 10 per cent.} and noise errors, and prior classifications in the literature.
    These allowed us to rule out another 22 candidates (13 SBPL, 9 PL).
\end{enumerate}
Even though some candidates could have been ruled out by multiple criteria above, our reported number of candidates ruled out by each criterion does not include those that were already eliminated by a previous criterion, i.e., this was a step-wise elimination process.
Our remaining sample contained five orphan afterglow candidates (listed in Table~\ref{tab:atca_observations}); of these, three were found from a SBPL fit and two from a PL fit. 

For each remaining candidate, we performed follow-up Australia Telescope Compact Array (ATCA) observations and present this in \textsection\ref{ssec:follow-up}.
In light of the new information obtained from these observations, we performed additional analysis to come to a final decision on whether each candidate is likely orphan afterglow related or can be explained by some other astrophysical phenomena.
We mainly arrived at these decisions for each candidate by checking (a) whether the light curve continued to exhibit a power-law behaviour at the epoch of the new ATCA observations (if not, whether interstellar scintillation would instead be a better alternative for explaining the observed variability in the light curve), and (b) whether the spectral/temporal evolution of the candidate was consistent with known GRB closure relations \citep[e.g.,][]{Granot2002} with a reasonable choice of microphysical parameters.
These details are presented and more comprehensively explained in \textsection\ref{ssec:candidates}.

\section{Results} \label{sec:results}
\subsection{Follow-up ATCA Observations} \label{ssec:follow-up}
We observed each candidate with the ATCA in order to better characterise their spectral properties and evolution. 
We conducted observations for all candidates on 2022 May 16 UT under project C3363 (PI: T. Murphy): at central frequencies of 2.1, 5.5, and 9.0\,GHz, each with 2\,048\,MHz bandwidth, and the array in 750D configuration.
The 750D configuration consisted of 10 short baselines $<800\,$m and 5 long baselines ${\sim} 4\,$km.
We also performed an additional epoch of observations for candidates VAST J195443.9$-$412511 and VAST J175036.1$-$181454.
For VAST J195443.9$-$412511, we also observed on 2022 September 8, at central frequencies of 2.1, 5.5, and 9.0\,GHz, each with 2\,048\,MHz bandwidth, and the array in 6D configuration (with a maximum baseline of 6\,km).
While for candidate VAST J175036.1$-$181454, we also observed on 2022 March 5, at central frequencies of 2.1, 5.5, 9.0, 16.7, and 21.2\,GHz, each with 2\,048\,MHz bandwidth, and the array in 6A configuration (with a maximum baseline of 6\,km). 
The observational and data reduction details are summarised in Table~\ref{tab:atca_observations}.

We reduced all the data using standard routines in {\sc Miriad} \citep{Sault1995}.
We used PKS 1934$-$638 to calibrate both the bandpass and the flux-density scale for all observations, with the exception of the $16.7/21.2$\,GHz observation for VAST J175036.1$-$181454, where we used PKS 1253$-$055 (3C 279) to calibrate the bandpass and PKS 1934$-$638 for the flux-density scale.
The calibrators we used to correct for the time-variable complex gains for each target source are listed in Table~\ref{tab:atca_observations}.
In the imaging process, we often flagged either the set of short or long baselines depending on the optimal ($u,v$)-coverage required for a reliable flux-density measurement.
This decision depended on various factors that were considered in an observation-by-observation basis, including hour-angle coverage, source elevation, observing frequency, required sensitivity and the extent of confusion from neighbouring sources (and their sidelobes).
We produced images using the multi-frequency synthesis CLEAN algorithm \citep{Hogbom1974, Clark1980, Sault1994}, mostly with a robustness of 0, and report the flux-density measurements in 
Tables~\ref{tab:sbpl_measurements1} and \ref{tab:pl_measurements1}.
The 2.1 to 9.0\,GHz spectral index $\alpha_{\rm 2.1-9\,GHz}$, obtained from an ordinary least-squares optimisation, is separately reported in Tables \ref{tab:sbplfits} and \ref{tab:plfits}.
We note that the data from the 21.2\,GHz observation for VAST J175036.1$-$181454 was adversely affected by poor weather conditions, while the observations of VAST J111757.5$+$021607 were affected by the very 1D ($u,v$)-coverage that results for observations of sources near the celestial equator with a linear East-West array -- as a result, no reliable measurements from these observations are reported here.

\begin{table*}
\centering
\caption{
Table of follow-up ATCA observations for the five orphan afterglow candidates.
Columns 1 through 6 show the candidate name, the right ascension and declination of the observation phase centre, the central frequency of the receiver, the gain calibrator used, and the baselines flagged in the imaging process.
All observations, with the exception of two, were taken on 2022 May 16, in the 750D array configuration, using PKS 1934$-$638 for both bandpass and flux-scale calibration.
The two exceptions are (a) VAST J195443.9$-$412511, which had an additional epoch of follow-up observations, taken on 2022 September 8 in the 6D array configuration; and (b) VAST J175036.1$-$181454, which had an additional epoch of follow-up observations, taken on 2022 March 5 in the 6A array configuration -- for only the 16.7\,GHz observation in this epoch, we used PKS 1253$-$055 (3C 279) to calibrate the bandpass and PKS 1934$-$638 for the flux-scale.
}
\label{tab:atca_observations}
\begin{tabular}{lccccl}
\hline\hline
Source Name &  
RA (J2000) &
Dec (J2000) &
Receiver (GHz) &
Gain Cal. &
Flagged Baselines \\
\hline
VAST J195443.9$-$412511 & 19:54:44.0 & $-41$:25:11.2 & 2.1 & 1954$-$388 & CA01--CA02 \\
&&& $5.5$ & \textquotedbl\textquotedbl & \textquotedbl\textquotedbl \\
(22 May 16) &&& $9.0$ & \textquotedbl\textquotedbl & CA01--CA02, all long baselines including CA06 \vspace{1mm} \\
\hdashline \vspace{-2.5mm} \\ 
(22 Sep 8) &&& 2.1 & \textquotedbl\textquotedbl & None \\
&&& $5.5$ & \textquotedbl\textquotedbl & \textquotedbl\textquotedbl \\
&&& $9.0$ & \textquotedbl\textquotedbl & \textquotedbl\textquotedbl \\
\hline
VAST J200430.5$-$401649 & 20:04:30.6 & $-40$:16:49.9 & 2.1 & 1954$-$388 & CA01--CA02 \\
&&& $5.5$ & \textquotedbl\textquotedbl & \textquotedbl\textquotedbl \\
&&& $9.0$ & \textquotedbl\textquotedbl & CA01--CA02, all long baselines including CA06 \\
\hline
VAST J054958.0$-$581946 & 05:49:58.0 & $-58$:19:46.4 & 2.1 & 0420$-$625 & CA01--CA02 \\
&&& $5.5$ & 0516$-$621 & all short baselines not including CA06 \\
&&& $9.0$ & \textquotedbl\textquotedbl & CA01--CA02, all long baselines including CA06 \\
\hline
VAST J111757.5$+$021607 & 11:17:57.5 & $+02$:16:07.3 & 2.1 & 1055$+$018 & all (not imaged due to 1D ($u,v$) coverage) \\ 
&&& $5.5$ & \textquotedbl\textquotedbl & \textquotedbl\textquotedbl \\ 
&&& $9.0$ & \textquotedbl\textquotedbl & \textquotedbl\textquotedbl \\ 
\hline
VAST J175036.1$-$181454 & 17:50:36.1 & $-18$:14:54.4 & 2.1 & 1730$-$130 & None \\
&&& $5.5$ & \textquotedbl\textquotedbl & \textquotedbl\textquotedbl \\
&&& $9.0$ & \textquotedbl\textquotedbl & \textquotedbl\textquotedbl \\
&&& $16.7$ & \textquotedbl\textquotedbl & \textquotedbl\textquotedbl \\
(22 Mar 5) &&& $21.2$ & \textquotedbl\textquotedbl & all (not imaged due to weather-related data issue) \vspace{1mm} \\
\hdashline \vspace{-2.5mm} \\ 
(22 May 16) &&& 2.1 & \textquotedbl\textquotedbl & all short baselines not including CA06 \\
&&& $5.5$ & \textquotedbl\textquotedbl & CA01--CA02, all long baselines including CA06 \\
&&& $9.0$ & \textquotedbl\textquotedbl & \textquotedbl\textquotedbl \\
\hline
\end{tabular}
\end{table*}

\subsection{Interpretation of Candidates} \label{ssec:candidates}
\begin{figure*}
    \includegraphics[width=0.86\linewidth]{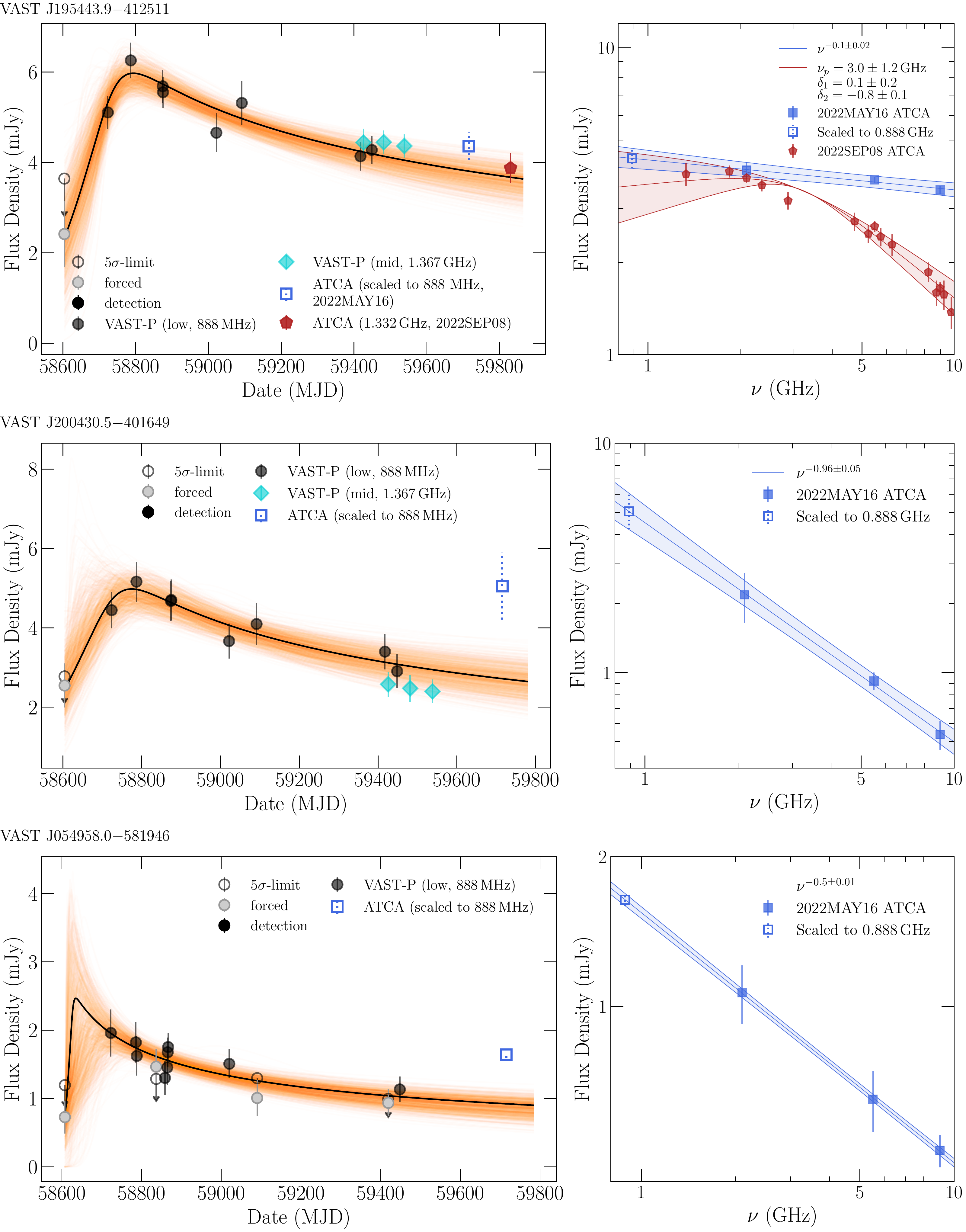}
	\vspace{-2mm}
    \caption{
    The radio light curves (\textit{left}) and spectra (\textit{right}) for each of the orphan afterglow candidates attained from a SBPL fit.
    For each light curve, only data from VAST-P and ATCA are shown; 
    the full radio dataset for each source, including other complementary data from archival radio surveys, can be found in Tables \ref{tab:sbpl_measurements1} and \ref{tab:pl_measurements1}.
    We represent the low-band (888\,MHz) detections with black circular markers, and for the non-detections, we show both a measurement from forced extraction (grey circular markers) and also a $5\sigma$ limit (open circular markers).
    The mid-band (1.367\,GHz) detections are shown with turquoise diamond markers (there are no mid-band non-detections). 
    The ATCA flux density measurements \textit{extrapolated} to 888\,MHz from the 2022 May 16 observations are shown with blue, open, square markers.
    For VAST J195443.9$-$412511, we show an additional ATCA data point taken from the 2022 September 8 observations -- this data point is represented by a red, filled, pentagonal marker and is from the 1.3\,GHz sub-band measurement, not an extrapolation.
    The black line is the fit to the light curve using parameters estimated from nested sampling (and overlaid are 1\,000 random samples to illustrate the fit uncertainties).
    For each radio spectrum, we extrapolated the ATCA measurements from 2022 May 16 (blue, filled, square markers) to 888\,MHz (blue, open, square markers) using a linear ordinary least-squares fit (blue line).
    For VAST J195443.9$-$412511, we show an additional radio spectral snapshot, using measurements taken from the 2022 September 8 ATCA observations (red pentagonal markers).
    For this 2022 September 8 radio spectrum, we fitted it with a smoothly broken power law (red line) with the estimated parameters for the peak frequency $\nu_\textrm{p}$, rise slope $\delta_1$, and decay slope $\delta_2$ given in the legend.
    }
    \label{fig:sbpl_lcsed}
\end{figure*}

\begin{figure*}
    \includegraphics[width=0.94\linewidth]{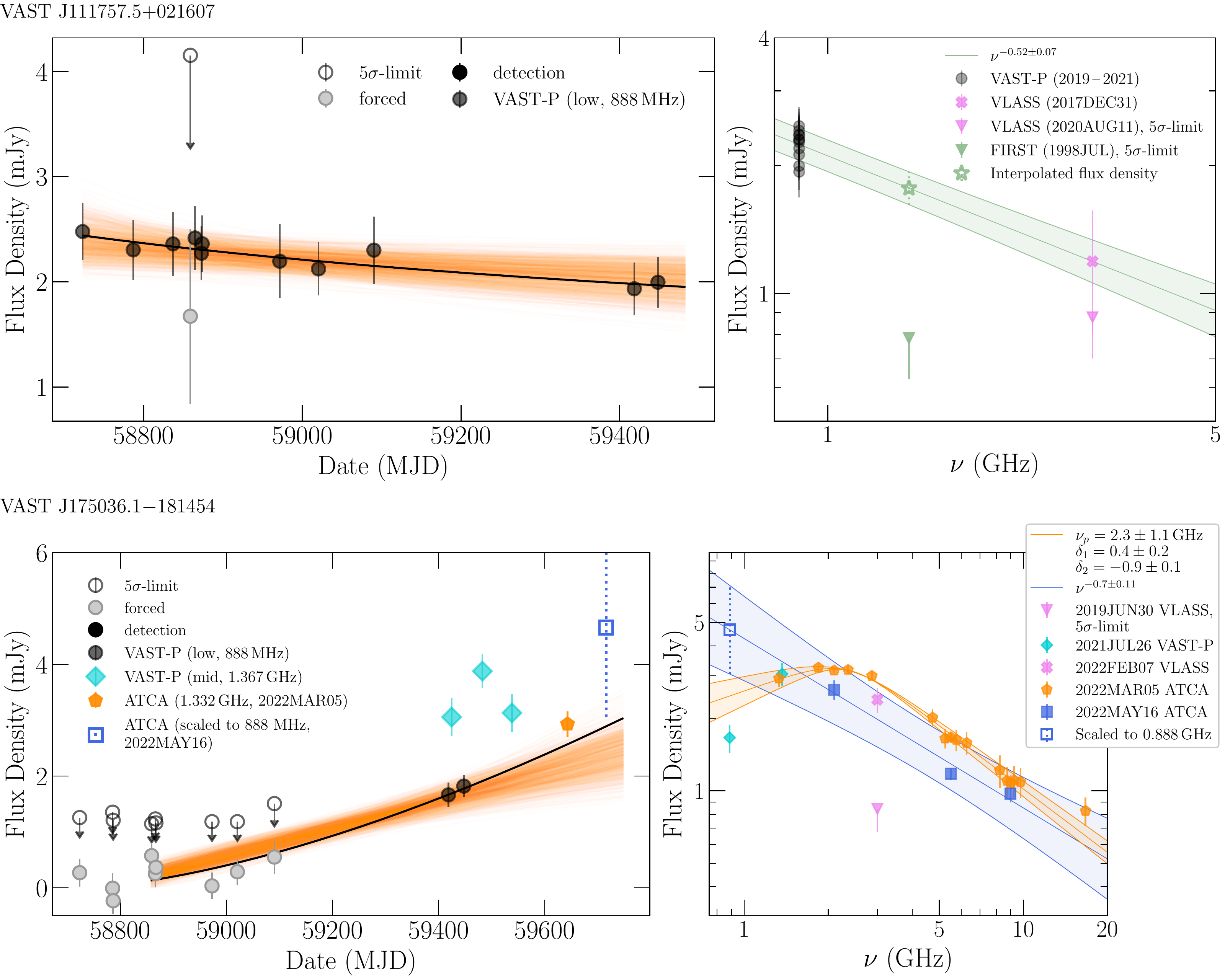}
	\vspace{0mm}
    \caption{
    The radio light curves (\textit{left}) and spectra (\textit{right}) for each of the orphan afterglow candidates attained from a PL fit, similar to Figure \ref{fig:sbpl_lcsed}. 
    For VAST J111757.5$+$021607, the radio spectrum consists of archival measurements from VAST-P (black circular markers), FIRST (green inverted triangular marker, indicating a non-detection $5\sigma$ limit) and VLASS (pink square marker for detection and pink inverted triangular marker to indicate a non-detection $5\sigma$ limit), instead of ATCA measurements.
    The VAST-P and VLASS detections were \textit{interpolated} to 1.4\,GHz (green, open, star marker) using a linear ordinary least-squares fit (green line).
    For VAST J175036.1$-$181454, the ATCA data points in the light curve include a $1.3$\,GHz sub-band measurement from the 2022 March 5 observation (orange, filled, pentagonal marker) and a flux density measurement \textit{extrapolated} to 888\,MHz from the 2022 May 16 observations (blue, open, square marker).
    We show the radio spectral snapshots from both ATCA observations for this candidate -- one from the 2022 March 5 epoch (orange pentagonal markers) and another from the 2022 May 16 epoch (blue square markers); we also show two VAST-P data points with turquoise diamond markers, and two VLASS data points with pink markers.
    The 2022 March 5 radio spectrum was fitted with a smoothly broken power law (orange line) with the estimated parameters for the peak frequency $\nu_\textrm{p}$, rise slope $\delta_1$, and decay slope $\delta_2$ given in the legend.
    The 2022 May 16 radio spectrum was fitted by an ordinary least-squares fit (blue line), which was used for extrapolating the 2022 May 16 flux density measurements to 888\,MHz (blue, open, square marker).
    }
    \label{fig:pl_lcsed}
\end{figure*}

For each of our five candidates, we repeated our functional fits to 
Equations~\ref{eq:sbpl} and \ref{eq:pl} to obtain more robust parameter (and error) estimates using the nested sampler {\sc Dynesty} \citep{Speagle2020} as implemented in the Bayesian inferences software {\sc Bilby} \citep{Ashton2019}.
We used the same physically motivated fit constraints from \textsection\ref{sec:search} with uniform priors, performing the nested sampling with 1\,000 live points and a stopping criterion on the estimated evidence $\hat{\mathcal{Z}}$ of $\Delta \mathrm{ln}(\hat{\mathcal{Z}}) = 0.05$.
The resulting parameter estimates are also reported in 
Tables~\ref{tab:sbplfits} and \ref{tab:plfits},
while the fitted light curves and radio spectra from ATCA observations are presented in Figures~\ref{fig:sbpl_lcsed} and \ref{fig:pl_lcsed}.

\begin{table*}
\caption{
The estimated temporal and spectral parameters for orphan afterglow candidates identified using a SBPL fit to the source light curve.
Column 1 is the candidate source name, Columns 2-6 show the median values (and 1$\sigma$ uncertainties) of the marginalised posterior distribution for each free parameter of the SBPL fit to the light curve attained from nested sampling, Column 7 is the spectral index from a least-squared fit to the ATCA measurements obtained on 2022 May 16.
}
\label{tab:sbplfits}
\begin{tabular}{lccccc|c}
\hline\hline
Source Name & $S_{{\rm peak}}$ & $\rm{{log}}(t_{{\rm peak}})$ & $\delta_1$ & $\delta_2$ & $\Delta t^{{1/3}}$ & $\alpha_{\rm 2.1-9\,GHz}$\\
\hline
VAST J195443.9$-$412511 & $5.63^{+0.26}_{-0.26}$ & $2.67^{+0.22}_{-0.25}$ & $2.96^{+2.96}_{-1.88}$ & $-0.47^{+0.12}_{-0.20}$ & $-6.92^{+2.29}_{-1.71}$ & $-0.10 \pm 0.02$ \\
VAST J200430.5$-$401649 & $4.79^{+0.33}_{-0.30}$ & $2.65^{+0.23}_{-0.25}$ & $2.43^{+3.04}_{-1.58}$ & $-0.61^{+0.20}_{-0.29}$ & $-6.86^{+2.14}_{-1.74}$ & $-0.96 \pm 0.05$ \\
VAST J054958.0$-$581946 & $1.99^{+0.74}_{-0.28}$ & $2.20^{+0.40}_{-0.51}$ & $3.77^{+3.75}_{-2.46}$ & $-0.47^{+0.12}_{-0.25}$ & $-4.60^{+2.01}_{-2.29}$ & $-0.50 \pm 0.01$ \\
\hline
\end{tabular}
\end{table*}

\begin{table*}
\caption{
The estimated temporal and spectral parameters for orphan afterglow candidates identified using a PL fit to the source light curve, similar to Table \ref{tab:sbplfits}.
There is no fitted spectral index for VAST J111757.5$+$021607 as there were no reliable measurements from ATCA for this source.
}
\label{tab:plfits}
\begin{tabular}{lccc|c}
\hline\hline
Source Name & $A$ & $\beta$ & $\Delta t^{{1/3}}$ & $\alpha_{\rm 2.1-9\,GHz}$\\
\hline
VAST J111757.5$+$021607 & $41.31^{+35.89}_{-26.80}$ & $-0.38^{+0.12}_{-0.08}$ & $-11.41^{+2.78}_{-4.28}$ & --- \\
VAST J175036.1$-$181454 & $0.00^{+0.01}_{-0.00}$ & $1.02^{+0.21}_{-0.21}$ & $2.54^{+1.82}_{-1.30}$ & $-0.55 \pm 0.08$ \\
\hline
\end{tabular}
\end{table*}

For each candidate, we also checked for counterparts in \textit{WISE}, and where possible, obtained a classification from the infrared colours (using the same classification regions as described in Section \ref{sec:search}, see Figure \ref{fig:wise_cc}).
We also compared the variability of each source to the expected extrinsic variability caused by refractive interstellar scintillation (RISS).
This effect is caused by the propagation of radio waves through the ionised interstellar medium in our Galaxy, causing the wavefront to be distorted, leading to phase changes resulting in the observed variability.
We used the NE2001 electron density model\footnote{We used the model via a python wrapper implemented in the package: \url{https://pypi.org/project/pyne2001/}} \citep{Cordes2002} to estimate the Galactic contribution to the electron distribution along the line of sight to our sources.
More specifically, the model outputs the transitional frequency $\nu_0$, the characteristic frequency where the transition from the strong to weak scattering scintillation regimes occur, and we used this to calculate the modulation index due to RISS for each source by applying the equation presented in \citet{Walker1998}: $V = (\nu/\nu_0)^{17/30}$, where $\nu$ is the observing frequency.
In the scenario where the source 
size $\theta_\textrm{S}$ exceeds the angular broadening diameter $\theta_\textrm{d}$ (which is dependent on the observing frequency and line of sight scattering measure), the modulation index (and degree of scintillation) would be reduced by a factor of $(\theta_\textrm{d} / \theta_\textrm{S})^{7/6}$.
We used this RISS variability analysis together with the light curves, radio spectra, and \textit{WISE} colour classifications to analyse each source.

\begin{figure*}
	\includegraphics[width=\linewidth,clip,trim={45mm 30mm 80mm 40mm}]{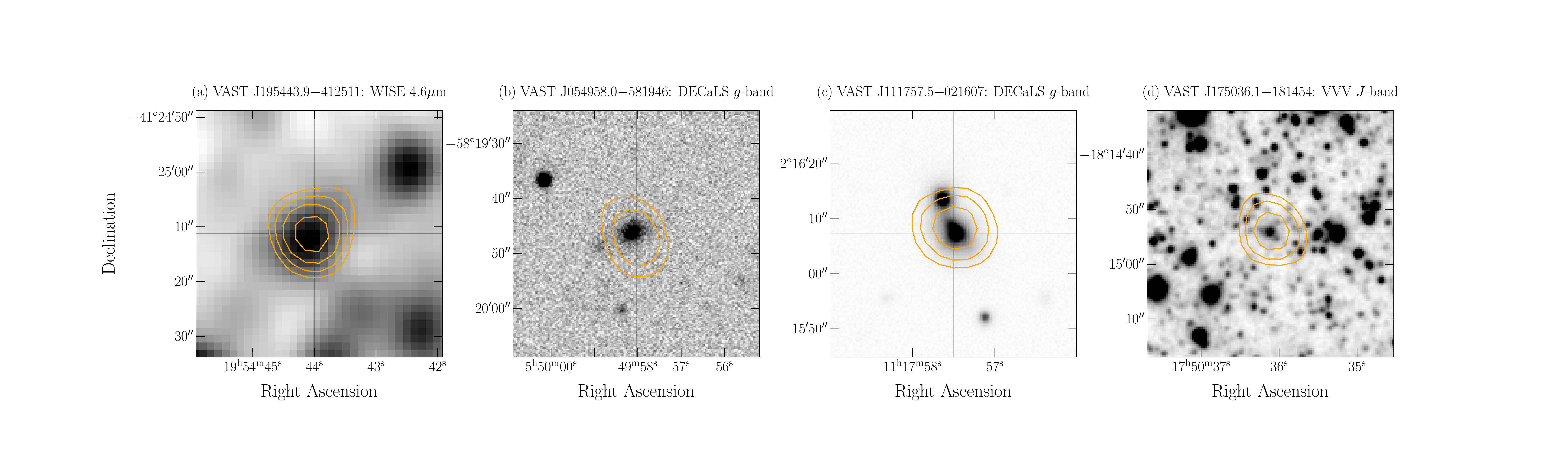}
    \caption{
    Images of multi-wavelength counterparts to the orphan afterglow candidates.
    Only candidates with a multi-wavelength counterpart are shown, with the image for each candidate selected from the highest resolution multi-wavelength survey available where a counterpart is detected.
    Presented from left to right are images of VAST J195443.9$-$412511 from the Wide-field Infrared Survey Explorer (WISE), VAST J054958.0$-$581946 from the Dark Energy Camera Legacy Survey (DECaLS), VAST J111757.5$+$021607 from DECaLS and VAST J175036.1$-$181454 from the VISTA Variables in the Via Lactea Survey (VVV).
    Overlaid are radio contours from the epoch of VAST with the highest signal-to-noise detection. 
    The radio contours increase by a factor of $\sqrt{2}$ at each step, with the lowest contour starting at the $3\sigma$ level.
    Each image is $45\arcsec \times 45\arcsec$, centred on the weighted average radio position across the VAST epochs, with North up and East to the left. 
    }
    \label{fig:mw_overlays}
\end{figure*}

\subsubsection{VAST J195443.9$-$412511}
VAST J195443.9$-$412511 was selected as a candidate via fitting to a SBPL; the fitted light curve (Figure \ref{fig:sbpl_lcsed}, \textit{upper-left panel}) rises with a slope of $2.96^{+2.96}_{-1.88}$ and decays with a slope of $-0.47^{+0.12}_{-0.20}$.
Our follow-up ATCA observations on 2022 May 16 show a flat spectrum extending across the 2.1 to 9\,GHz frequency range (see Figure \ref{fig:sbpl_lcsed}, \textit{upper-right panel}), with a spectral index of $\alpha_{\rm 2.1-9\,GHz} = -0.10 \pm 0.02$.
This spectral index is more shallow than any possible spectral segment expected for afterglow emission and cannot be explained by the expected curvature of the respective spectral breaks \citep[e.g.,][]{Granot2002} so we rule out the possibility that this source is afterglow related.
We obtained better spectral resolution in a subsequent epoch of ATCA observations taken on 2022 September 8 by splitting 
each observing band into four sub-bands, each with a bandwidth of 512\,MHz, prior to imaging (see Table~\ref{tab:sbpl_measurements1}).
These observations revealed
a steepening of the spectrum, particularly at higher frequencies, and the presence of spectral curvature; a fit of this radio spectrum to a SBPL located a spectral turnover at $\nu = 3.0 \pm 1.2$\,GHz with a rise and decay index of $\delta_1 = 0.1 \pm 0.2$ and $\delta_2 = -0.8 \pm 0.1$, respectively.

The flat radio spectrum observed in the first epoch of ATCA observations likely indicates the presence of a compact emitting region in a radio galaxy or an AGN jet aligned with our line of sight \citep[i.e., a blazar;][]{Urry1995}.
The variability from the compact emitting region could either be intrinsic or extrinsic to the source.
In the intrinsic scenario, we used light-crossing time arguments to constrain the size of the emitting region $r$ to $r < \tau c = 0.16\,$pc, where $c$ is the speed of light and $\tau$ is the timescale of variability, estimated as the time between the minimum and maximum flux measurement in our light curve (Figure \ref{fig:sbpl_lcsed}, \textit{upper-left panel}).
While this limit is broadly consistent with the inferred size of jets emanating from supermassive black holes detailed in the literature \citep[e.g.,][]{Ighina2022}, this upper-limit estimate does not factor in the possibility that variability could travel quicker than the light-crossing time in the scenario of relativistic boosting 
and apparent superluminal motion.
Alternatively, if the variability is attributed to extrinsic RISS, the observed modulation index is consistent with the expected modulation index from RISS; i.e., we can also put an angular constraint on the emission region to be smaller than the angular broadening diameter in the source's line of sight $\theta_\textrm{d} = 1.4\,$mas.
Although the rapid rise and prolonged decay could be produced by a scattering screen with somewhat constrained properties, we feel that the intrinsic scenario is more likely to be correct.

The observed properties of VAST J195443.9$-$412511 -- the rapid rise in the light curve and prolonged decay, the flat-to-steep radio spectrum transition, the high-to-low frequency evolution, and the evolution timescale of the source on the order of months -- are similar to those observed in quasar 3C273 flares \citep[e.g.,][]{Robson1983}.
The observed properties of such flares have been explained by models invoking relativistic shocks originating from disturbances in the outflow of an adiabatic, conical jet \citep[e.g.][]{Marscher1985,ODell1988}.
One caveat to this interpretation is the detection of a \textit{WISE} counterpart (see Figure~\ref{fig:mw_overlays}, \textit{panel a}); although the non-detection in the 12\,\textmu m band prevents a conclusive source classification from using only \textit{WISE} colours 
($[3.4\,\textrm{\textmu m}] - [4.6\,\textrm{\textmu m}] = 0.15 \pm 0.09$\,mag and $[4.6\,\textrm{\textmu m}] - [12\,\textrm{\textmu m}] < 3.04$\,mag),
the colours provide sufficient constraints to disfavour an AGN origin (see Figure~\ref{fig:wise_cc}).
Alternatively, for interpreting the observed properties of VAST J195443.9$-$412511, we have also considered the possibilities of a jet quenching in X-ray binaries \citep[e.g.,][]{Russell2020} and radio flares from TDEs \citep[e.g.,][]{Horesh2021}.
However, we rule out these since the former is expected to evolve on timescales much shorter on the order of hours and days (even though it can explain the flat-to-steep radio spectrum transition), while the latter struggles to explain the initially observed flat spectrum (even though it can explain the light-curve evolution).
While we currently consider the AGN flare to be the leading interpretation, a definitive source classification would require spectroscopic follow-up and a comprehensive characterisation would benefit 
from continued multi-frequency monitoring to trace the evolution of the source's radio spectra as well as VLBI follow-up to constrain the source size.

\subsubsection{VAST J200430.5$-$401649}
VAST J200430.5$-$401649 was selected as a candidate via fitting to a SBPL; the fitted light curve (Figure \ref{fig:sbpl_lcsed}, \textit{middle-left panel}) rises with a slope of $2.43^{+3.04}_{-1.58}$ and decays with a slope of $-0.61^{+0.20}_{-0.29}$.
The source was not detected in archival radio surveys and has no known counterparts in any multi-wavelength survey.
ATCA follow-up revealed that it had a spectral index of $\alpha_{\rm 2.1-9\,GHz} = -0.96 \pm 0.02$, consistent with the spectral index of $\alpha \sim -1$ inferred from the low-band epoch 17 and mid-band epoch 18 measurements taken one week apart.
By extrapolating the radio spectrum from the ATCA observations (Figure \ref{fig:sbpl_lcsed}, \textit{middle-right panel}), the scaled flux density at 888\,MHz on 2022 May 16 is $5.1 \pm 0.9$\,mJy.
This is a significant deviation away from the temporal power-law decay expected from an orphan afterglow (see Figure \ref{fig:sbpl_lcsed}, \textit{middle-left panel}); we therefore disfavour an afterglow interpretation for this variable source.
The source has a modulation index of $V = 0.22$ (calculated from the low-band VAST-P data points), consistent with the expected variability from RISS along this line of sight.

In the analysis above where we disfavour the afterglow scenario, we note that extending the single power-law component from the ATCA frequencies to 888\,MHz requires us to assume the absence of a spectral break at and around these frequencies.
This assumption is at odds with the radio spectra of VAST J195443.9$-$412511 and VAST J111757.5$+$021607 in Figures~\ref{fig:sbpl_lcsed} and \ref{fig:pl_lcsed}, respectively; these plots instead show that the extrapolations have a tendency to lead to an overprediction due to the possible presence of spectral curvature.
Despite this, we reason below that for the sole purpose of disfavouring the afterglow scenario, where we can use the fireball model to describe the location of the spectral turnover, it is valid to assume the absence of a spectral break for this extrapolation (and this analysis).
This is a reasonable assumption because the injection break $\nu_\textrm{m}$ is expected to be $\ll 888$\,MHz at the epoch of our ATCA observation based on an estimation using \textit{conservative} afterglow parameters in the following relation described in \citet{Panaitescu2000}:
\begin{equation}
    \nu_\textrm{m} = 1.9 \times 10^{13} \, E^{1/2}_{\textrm{iso,kin},53} \, \epsilon^{2}_{\textrm{e},-1} \, \epsilon^{1/2}_{\textrm{B},-2} \, T_\textrm{d}^{-3/2} \, (1+z)^{1/2} \, \textrm{Hz},
\end{equation}
where $E_\textrm{iso,kin} = 10^{53} E_{\textrm{iso,kin},53}\,$erg is the total isotropic equivalent kinetic energy of the blast wave, 
$\epsilon_\textrm{e} = 0.1 \epsilon_{\textrm{e},-1}$ and $\epsilon_\textrm{B} = 0.01 \epsilon_{\textrm{B},-2}$ are the fractional shock energies in the accelerated electrons and magnetic fields, respectively,
$z$ is the redshift to the source, and
$T_\textrm{d}$ is the number of days elapsed since the burst event.
In particular, this equation assumes a uniform jet, a wind environment\footnote{We selected the equation in \citet{Panaitescu2000} that described a wind environment since this was the more conservative option, producing a $\nu_\textrm{m}$ value that was a factor of ${\sim} 2$ higher than that in a homogeneous interstellar medium environment.}, no additional energy injections, and an electron spectral index $p=2.5$, with the conservative input parameters we used taking values within a physically reasonable range that maximise $\nu_\textrm{m}$.
Specifically, the input parameters we used were: 
\begin{itemize}
    \item $E_\textrm{iso,kin} = 10^{54}$\,erg -- most bursts have $E_\textrm{iso,kin} \sim 10^{53}\,$erg, but the GRBs at the energetic end of the distribution will have $E_\textrm{iso,kin}$ values an order of magnitude higher \citep[e.g.,][]{Panaitescu2001,Cenko2011};
    \item $\epsilon_\textrm{e} = 0.1$ -- the distribution of $\epsilon_\textrm{e}$ is narrowly around ${\sim} 0.1$ \citep[e.g.,][]{Beniamini2017};
    \item $\epsilon_\textrm{B} = 10^{-5}$ -- a systematic study of magnetic fields in GRB shocks showed that in a wind environment the range for this parameter is from $10^{-7}$ to $10^{-5}$ \citep[][]{Santana2014}; 
    \item $T_\textrm{d} = 1\,100\,$days -- this chosen value corresponds to the time between the epoch of the first ASKAP detection and the epoch of the ATCA observations, for which we are performing the extrapolation, i.e., the minimum possible onset time for the GRB event; and
    \item $z = 0.2$ -- the limiting redshift which we expect an afterglow to be detectable in our surveys.
\end{itemize}
These conservative input parameters predicts the location of the injection frequency at the time of the ATCA observations to be $\lessapprox 200\,$MHz.
However, owing to the significant uncertainty in the range of values $\epsilon_\textrm{B}$ can take (spanning many orders of magnitude in the literature), we only disfavour rather than completely rule out an afterglow origin when considering candidates using this method.
VAST J054958.0$-$581946 was the only other source where we applied this method (with the same set of assumptions and reasoning) to disfavour the afterglow scenario. 

\subsubsection{VAST J054958.0$-$581946}
VAST J054958.0$-$581946 was selected as a candidate via fitting to a SBPL; the fitted light curve (Figure \ref{fig:sbpl_lcsed}, \textit{lower-left panel}) rises with a slope of $3.77^{+3.75}_{-2.46}$ and decays with a slope of $-0.47^{+0.12}_{-0.25}$.
The source is associated with both \textit{WISE} and DECaLS counterparts (see Figure \ref{fig:mw_overlays}, \textit{panel b}); the \textit{WISE} colours 
($[3.4\,\textrm{\textmu m}] - [4.6\,\textrm{\textmu m}] = 0.33 \pm 0.05$\,mag and $[4.6\,\textrm{\textmu m}] - [12\,\textrm{\textmu m}] = 2.28 \pm 0.30$\,mag)
suggest the source is associated with a spiral galaxy at a photometric redshift (as determined by DECaLS) of $z_\textrm{photo} = 0.218$.
ATCA follow-up revealed that it had a spectral index of $\alpha_{\rm 2.1-9\,GHz} = -0.50 \pm 0.01$.
By extrapolating the radio spectrum from the ATCA observations (Figure \ref{fig:sbpl_lcsed}, \textit{lower-right panel}), the scaled flux density at 888\,MHz on 2022 May 16 is $1.64 \pm 0.05$\,mJy.
This is a significant deviation away from the temporal power-law decay expected from an orphan afterglow (see Figure \ref{fig:sbpl_lcsed}, \textit{lower-left panel}); 
we therefore disfavour the possibility the radio emission is afterglow related (following a similar argument to VAST J200430.5$-$401649) and propose it is most likely associated with the spiral galaxy.
The source has a modulation index of $V = 0.26$ (calculated from low-band VAST-P data points), consistent with the expected variability from RISS.

\subsubsection{VAST J111757.5$+$021607}
\label{ssec:candidate4}
VAST J111757.5$+$021607 was selected as a candidate via fitting to a decaying PL; the fitted light curve (Figure \ref{fig:pl_lcsed}, \textit{upper-left panel}) decays with a shallow slope of $-0.38^{+0.12}_{-0.08}$.
The source is associated with both \textit{WISE} and DECaLS counterparts (see Figure \ref{fig:mw_overlays}, \textit{panel c}); the \textit{WISE} colours 
($[3.4\,\textrm{\textmu m}] - [4.6\,\textrm{\textmu m}] = 0.34 \pm 0.04$\,mag and $[4.6\,\textrm{\textmu m}] - [12\,\textrm{\textmu m}] = 4.09 \pm 0.05$\,mag)
suggest the source is associated with a starburst galaxy at a photometric redshift (as determined by DECaLS) of $z_\textrm{photo} = 0.079$, i.e., at a luminosity distance of 370\,Mpc.

Even though ATCA follow-up observations could not provide useful measurements due to the poor ($u,v$)-coverage arising from the source's proximity to the celestial equator, reliable measurements of the source at a higher frequency were obtained from quick-look images from two separate VLASS epochs taken on 2017 Dec 31 and 2020 Aug 11, respectively.
The source was detected in the first VLASS epoch at $1.19 \pm 0.38$\,mJy \citep[measurement from the VLASS Quick Look Images catalogue;][]{Gordon2021}, but was below the $5\sigma$ detection threshold of $0.88$\,mJy in the second epoch.
Since the uncertainties from the detection in the first epoch overlaps with the $5\sigma$ limit in the second epoch, we could not assess the degree nor significance of variability for the source at 3\,GHz between the two epochs. 
The VAST-P measurements alone, however, indicate little variability\footnote{Despite the small amount of variability, the source passed our $F$-statistic criterion, which compared the PL model to the constant benchmark model.
This is because, while both models explained the data quite well with $\chi^2_\textrm{PL} = 1.24$ and $\chi^2_\textrm{benchmark} = 4.74$ (corresponding to $p$-values of 0.998 and 0.943, respectively), the $F$-statistic determines whether the PL fit is \textit{significantly} better than the benchmark based on the ratio of the $\chi^2$ values, not the absolute difference (see Equation~\ref{eq:F}).
The resulting $F = 12.77$ corresponds to a $p$-value of 0.002 under the $F$-distribution with $(2,9)$ degrees of freedom, indicating a \textit{significantly} better fit for the PL model over the benchmark model.}, with the source having a modulation index of $V = 0.11$ and reduced $\chi^2$ of $\eta = 0.43$ (calculated from the low-band VAST-P data points).

Taking these results, we assumed the source had limited variability and performed a spectral fit between the detections from
the VAST-P observations taken at 888\,MHz and the VLASS epoch 1 observation taken 2\,yr prior at 3\,GHz.
This yielded a spectral index of $-0.52 \pm 0.07$; 
interpolating the radio spectrum to 1.4\,GHz gives a flux density of $1.77 \pm 0.15$\,mJy, which is inconsistent with FIRST observations taken in 1998 July as shown in Figure \ref{fig:pl_lcsed} (\textit{upper-right panel}).
Those FIRST observations yielded a non-detection with a $5\sigma$ limit of 0.78\,mJy, though there is a spatially coincident peak $4\sigma$ above the noise floor at ${\sim} 0.6$\,mJy.
This suggests our prior assumption that the source has limited variability does not hold and instead indicates source brightening of \textit{up to}\footnote{This is an upper limit since we must use the $5\sigma$ limit from the second epoch of VLASS, taken at a similar time as the VAST-P epochs, for the interpolation given that the source flux density evolves over time. The interpolation of this non-detection limit to 1.4\,GHz yields an upper limit of 1.6\,mJy.} a factor of ${\sim} 3$ in the 20\,yr since the FIRST observation.

A possible interpretation of these observations could be a variable or transient with variability on decade-long timescales found in a starburst radio galaxy.
If this source is related to an orphan afterglow, it would decay similarly to the afterglow from that of a standard on-axis GRB (e.g., see figure 1 in \citealt{Granot2018a}); the fitted slope of $-0.38^{+0.12}_{-0.08}$ suggests an extreme, but possible \citep[e.g.,][]{Ghisellini2009, Wang2015}, electron spectral index $p {\sim} 1.5$ for a burst exploding in a homogeneous interstellar-medium environment in the slow-cooling regime \citep[e.g.,][]{Sari1998}, following the relationship $S \propto t^{3(1-p)/4}$.
However, these closure relations also necessitate, in this scenario, the spectrum to follow $S \propto \nu^{(1-p)/2}$; a value for $p {\sim} 1.5$ implies a spectral index of $\alpha \sim -0.25$, which is inconsistent with the spectral index of $-0.52 \pm 0.07$ fitted earlier using the epoch 1 VLASS detection (if we instead considered the more temporally aligned epoch 2 VLASS upper limit, the spectral index is even steeper, $\alpha < -0.8$, allowing us to arrive at the same conclusion).
We therefore rule out the possibility the radio emission is afterglow related, but do not rule out the possibility it is produced by a non-afterglow related transient with variability on decade-long timescales.

In an alternate scenario treating the FIRST observations as an aberration, the low variability of the source with a modulation index smaller than that expected from RISS by a factor of three indicates the primary contribution to the radio emission is extended, emanating from the star-forming galaxy.
The detection in the higher resolution VLASS epoch 1 data also supports the interpretation of extended radio emission, with an integrated-to-peak flux-density ratio of $S_I/S_P = 1.8$, though there are some systematic uncertainties affecting the integrated flux density differently to the peak flux density particularly at low flux densities \citep{Lacy2020,Gordon2021}.
Under this interpretation, the radio emission may have been resolved out to a degree in the FIRST observations, which have comparable resolution to that of VLASS.
The extended radio emission, arising from synchrotron processes associated with relativistic electrons accelerated in supernova remnants, could be used to infer the star formation rate (SFR) from recent star formation ($10^8$\,yr) occurring in the galaxy \citep{Condon1992}.
To calculate the SFR, we applied the relationship presented in \citet{Greiner2016}:
\begin{equation}
    \frac{\textrm{SFR}}{M_\odot\,\textrm{yr}^{-1}} = 0.059\,\bigg(\frac{S_\nu}{\textrm{\textmu Jy}} \bigg)\,(1+z)^{-(\alpha+1)}\,\bigg(\frac{D_\text{L}}{\text{Gpc}}\bigg)^2\,\bigg(\frac{\nu}{\text{GHz}}\bigg)^{-\alpha},
	\label{eq:sfr-radio}
\end{equation}
where $S_\nu$ is the total flux density, $z$ is the redshift, $\alpha$ is the spectral index of the radio emission, $D_{\text{L}}$ is the luminosity distance and $\nu$ is the observing frequency.
The radio-derived SFR is $9.37\,M_\odot\,\textrm{yr}^{-1}$ and this is consistent with previous studies of radio-derived SFR from star-forming radio galaxies in the local Universe \citep[e.g.,][]{Bonzini2015}.
Although we currently favour this interpretation, continued long-term monitoring on the timescale of years would enable us to confirm whether the radio emission from VAST J111757.5$+$021607 is due to star formation from the host galaxy or the presence of a slowly evolving radio transient.

\subsubsection{VAST J175036.1$-$181454}
\label{ssec:candidate5}
VAST J175036.1$-$181454 was selected as a candidate via fitting to a rising PL; the fitted light curve (Figure \ref{fig:pl_lcsed}, \textit{lower-left panel}) rises with a slope of $1.02^{+0.21}_{-0.21}$.
This candidate is the only candidate in our sample we do not rule out as possibly afterglow related.

The radio spectrum inferred from the quasi-simultaneous low-band epoch 17 and mid-band epoch 18 observations taken one week apart appears to be inverted (positive), 
in contrast to the higher frequency ATCA observations which had a negative spectral slope.
For each of the 2.1, 5.5 and 9.0\,GHz frequency bands in the 22 March 5 ATCA observations, we split the 2\,048\,MHz bandwidth into four sub-bands, each with a bandwidth of 512\,MHz, and imaged them to obtain a better spectral resolution (see Table~\ref{tab:pl_measurements1}).
The resulting radio spectrum shown in Figure \ref{fig:pl_lcsed} (\textit{lower-right panel}) reveals a spectral turnover at $\nu = 2.3 \pm 1.1$\,GHz with a rise and decay index of $\delta_1 = 0.4 \pm 0.2$ and $\delta_2 = -0.9 \pm 0.1$, respectively, obtained from a fit to a SBPL.
Due to a range of different factors in the 22 May 16 observations, including shorter integration time, more significant radio frequency interference, and poorer ($u,v$)-coverage, we could not obtain reliable flux-density measurements for the sub-bands to check whether the location of the spectral turnover had shifted; however, we note that the decay slope of $\alpha_{\rm 2.1-9\,GHz} =  -0.7 \pm 0.1$ is broadly consistent with the previous epoch of ATCA observations and the flux density faded by a factor of ${\sim} 1.3$ across the entire radio spectrum.

The first epoch of ATCA observations also showed evidence for the source being extended at smaller spatial scales: while it was unresolved at 2.1\,GHz with a beam size of $14.05\arcsec \times 2.47\arcsec$, it had an integrated-to-peak flux-density ratio of $S_I / S_P \sim 1.4$ in each of the 5.5, 9.0, and 16.7\,GHz observing bands, with beam sizes of $7.19\arcsec\times1.01\arcsec$, $4.44\arcsec\times0.65\arcsec$, and $3.73\arcsec\times0.44\arcsec$, respectively.
This suggests that if the radio source is a transient, it could have some low-level contamination from its host galaxy or a slightly-extended foreground source.
However, the source was unresolved in all bands of the second epoch of ATCA observations, which had lower angular resolution due to shorter baselines in the array configuration, as well as in the 2022 February 7 VLASS observation with beam size $2.82\arcsec\times1.67\arcsec$ (i.e., with angular resolution comparable to those in the first epoch of ATCA observations).

Since the source was not detected prior to epoch 17 (i.e., no detections in VAST-P1 or archival surveys), we classified this source as a transient.
The modulation index of the source (calculated from the low-band VAST-P data points) $V = 1.29$ (and a reduced $\chi^2$ of $\eta = 9.13$) cannot be explained by extrinsic variability from RISS (expected to be $V \sim 0.09$), which supports the classification of the source as a transient. 
It is located close to the Galactic plane ($l = 9.78\degr$, $b = 4.53\degr$) and $10.8\degr$ away from the Galactic Centre, though we rule out the possibility of this source being a Galactic Centre Radio Transient \citep[this was a transient type we previously found near the Galactic Centre in a VAST-P1 variability search;][]{Wang2021} since it does not exhibit a steep spectrum nor show evidence of Stokes \textit{V} emission in any of our radio datasets.

If we consider the source under the assumption its temporal and spectral evolution can be described by synchrotron radiation, which is common for extragalactic transients, we can independently estimate the electron spectral index $p$ from the light-curve decay, using the relation $S \propto t^{(1-3p)/4}$, and the optically thin radio spectrum, using the relation $S \propto \nu^{(1-p)/2}$ \citep[e.g.,][]{Granot2002}. 
The analysis and assumptions used here are similar to what we have outlined in \textsection\ref{ssec:candidate4} when we assessed the viability of an orphan afterglow interpretation for VAST J111757.5$+$021607; the main difference here is that we have used relations valid for a burst exploding into a stellar wind environment (typical of long GRB and supernova environments) instead of an homogeneous interstellar medium environment (typical of short GRB environments) since we later show the radio luminosity of this transient disfavours a short GRB origin. 
Assuming the transient occurred approximately halfway between the last epoch of non-detection and the first epoch of detection, the temporal decay index inferred from the $5.5$\,GHz\footnote{We did not consider 2.1 and 9\,GHz observations here because their reliability for estimating $p$ is affected by the spectral turnover and low signal-to-noise, respectively.} ATCA observations is $\beta = -2.2 \pm 0.6$, corresponding to an electron spectral index of $p = 3.2 \pm 0.8$.
We note that while the uncertainties are large since there are only two data points available after the light-curve turnover (and the uncertainties on the ATCA measurements are larger after accounting for systematics), this value is consistent with the more reliable estimate inferred from the optically thin segment of the high-signal radio spectrum obtained from the the first epoch of ATCA observation, $p = 2.8 \pm 0.2$.
This inferred value for $p$ is broadly consistent with the higher end of the electron spectral index distribution expected for GRBs \citep[e.g.,][]{Wang2015}, but also for other synchrotron transients.

A near-infrared counterpart with an association separation of ${\sim} 0\farcs4$ was found in the VISTA Variables in the Via Lactea Survey (VVV) DR2 catalogue \citep[][see Figure \ref{fig:mw_overlays}, \textit{panel d}]{Minniti2010,Minniti2017}, where it was morphologically classified as a galaxy.
An optical counterpart with an association separation of ${\sim} 0\farcs3$ was also detected in a Pan-STARRS sky survey, allowing for the determination of a photometric redshift $z_\textrm{photo} = 0.25 \pm 0.02$ \citep{Beck2021}.
With a chance coincidence probability\footnote{Calculated as $p_\textrm{cc} = 1 - \textrm{exp}(-\pi r^2 \sigma_{\leq m})$, where $r$ is the angular offset of the counterpart from the radio source location and $\sigma_{\leq m}$ is the average surface density of sources brighter than the counterpart magnitude $m$ in the region near the source.} of 0.6 per cent and 0.7 per cent respectively, we interpret these counterparts as the candidate host galaxy associated with the transient.

At the distance inferred from the photometric redshift, the transient has a measured peak spectral luminosity of $L_\nu = (7.5 \pm 0.6) \times 10^{30}$\,erg\,s$^{-1}$\,Hz$^{-1}$ at 1.4\,GHz and $\nu L_\nu = (1.1 \pm 0.1) \times 10^{40}$\,erg\,s$^{-1}$, where the errors are propagated from the statistical uncertainties of the flux density measurements only. 
The inferred peak luminosity of this transient is incompatible with most classes of transients, including fast luminous transients\footnote{One exception to this is ZTF18abvkwla (the ``Koala''), which had also reached $\nu L_\nu \gtrapprox 10^{40}$\,erg\,s$^{-1}$ -- this is more than an order of magnitude higher than other fast luminous transients in the literature. The rise time of the ``koala'' at radio frequencies, in particular, occurred on shorter timescales than what was observed for VAST J175036.1$-$181454. For these reasons, we currently disfavour a fast luminous transient origin for this transient, but do not rule it out entirely -- a more careful consideration will be discussed in future work.}, short GRBs and all classes of supernovae \citep[e.g., see figure 9 in][]{Ho2020b}; only the most luminous classes of transients, i.e., long GRBs and the sub-class of relativistic TDEs \citep[e.g., \textit{Swift} J164449.3$+$573451;][\citealt{Alexander2020}]{Zauderer2011}, are consistent with the inferred peak luminosity of VAST J175036.1$-$181454.

In the long GRB scenario, we used an updated version of the catalogue compiled in \citet{Leung2021} to check for any GRBs that are both temporally (occurring between the third last non-detection and the first detection) and spatially (GRB error region containing the transient sky direction) consistent with the transient; we found no corresponding high-energy trigger and therefore classify this transient as a strong orphan afterglow candidate.
The catalogue we used for these checks also included sub-threshold triggers from the \textit{Swift}/Burst Alert Telescope (BAT) Gamma-Ray Urgent Archiver for Novel Opportunities \citep[GUANO;][]{Tohuvavohu2020} programme, low-significance \textit{INTEGRAL} WEAK alert events\footnote{These are circulated via the \textit{INTEGRAL} Burst Alert System \citep{Mereghetti2003} and can be found at: \url{http://ibas.ncac.torun.pl/~jubork/ibas/ibas.php?slcn=weak}} and unclassified \textit{Fermi}/Gamma-ray Burst Monitor (GBM) triggers\footnote{These include \textit{all} triggers (including those not classified as GRBs) found in the following database maintained by NASA/GSFC: \url{https://heasarc.gsfc.nasa.gov/W3Browse/fermi/fermigtrig.html}}.
We also found no association with any known TDEs after checking with published databases\footnote{The databases we used included the Transient Name Server (maintained by the IAU supernova working group, \url{https://www.wis-tns.org/}) and the Open TDE Catalogue (maintained by James Guillochon, \url{https://tde.space})}, which was expected in the off-axis TDE scenario.

This transient is the target of an ongoing follow-up programme, involving the addition of low-frequency and high-resolution radio facilities, high-energy follow-up as well as spectroscopic observations.
The goal of this programme is to definitively classify and better characterise the transient; we will present the results from this programme in a future work.

\section{Discussion and Conclusions} \label{sec:d+c}
The follow-up observations and analysis of the five orphan afterglow candidates revealed the following:
\begin{enumerate}
    \item one candidate is likely a synchrotron transient, with an off-axis afterglow or an off-axis TDE as the leading interpretations; 
    \item one candidate is likely a flaring AGN, displaying a flat-to-steep radio-spectral transition over the span of 4\,months; 
    \item one candidate is associated with a starburst galaxy, with the radio emission originating from either star formation or an underlying slowly evolving transient; and 
    \item  the remaining two candidates are likely extrinsic variables caused by interstellar scintillation.    
\end{enumerate}
We discuss and summarise the possible implications of these results on transient and GRB rates as well as for future radio transient studies.

\subsection{Rates and the Inverse Beaming Fraction}
From this work, we found one extragalactic radio transient in the VAST-P, which repeatedly covered a footprint of ${\sim} 5\,000$\,deg$^{2}$ over a 2.5-yr span.
Considering ${\sim} 30\,$d transient timescales \citep[typical of gigahertz extragalatic synchrotron sources at our survey sensitivity; e.g.,][]{Ghirlanda2014,Metzger2015}, the effective sky area covered by VAST-P (low-band) is $35\,820\,$deg$^2$, implying a surface transient density of $(2.79^{+12.8}_{-2.72}) \times 10^{-5} \,$deg$^{-2}$ (the upper and lower limits represent the 95 per cent confidence interval as defined in \citealt{Gehrels1986}) at the flux-density threshold of ${\sim} 2\,$mJy\footnote{Our search criteria in \textsection\ref{sec:search} requires at least two $7.5\sigma$ detections or one $10\sigma$ detection. With $\sigma_\textrm{rms} = 0.24$\,mJy\,beam$^{-1}$, this is equivalent to ${\sim} 2$\,mJy.}.
We note that: (a) comparisons with previous variability surveys should be taken with care since many sources we had ruled out as being transients would have been classified as transients in previous surveys which spanned much shorter time periods, and (b) the rate can be considered a conservative lower limit rate as it does not account for possible detection of transients from traditional statistical search techniques (although our preliminary analysis of the VAST-P data suggests no additional radio transients have been found).

Estimating the event rate for objects similar to the one transient we had found, VAST J175036.1$-$181454, we find:
\begin{equation}\label{eq:rate}
\mathcal{R} \approx \frac{1\,\textrm{event}}{\Delta t \Delta \Omega (\frac{4}{3}\pi d^3)C_\textrm{2mJy}} \approx 0.52^{+2.36}_{-0.51}\,\textrm{Gpc}^{-3}\,\textrm{yr}^{-1},    
\end{equation}
where $\Delta t \sim 2.5\,$yr is the temporal baseline of VAST-P (low-band), $\Delta \Omega \sim 0.12$ is the fractional sky coverage of the VAST-P (low-band) footprint, $d \sim 1.3\,$Gpc is the maximum distance at which a source as luminous as VAST J175036.1$-$181454 would be detectable above ${\sim} 2\,$mJy, and ${C_\textrm{2mJy} \sim 0.7}$ is the completeness of VAST-P (low-band) at 2\,mJy \citep{Hale2021}.
The corresponding 95 per cent confidence interval is then 0.01-2.88\,$\textrm{Gpc}^{-3}\,\textrm{yr}^{-1}$.
This estimated rate for VAST J175036.1$-$181454-like objects is compatible with predicted rates for off-axis long GRBs and off-axis TDEs, but are too small to be consistent with short GRBs and supernovae; this is in agreement with the surface transient density inferred from the search results, which is also compatible with predictions for off-axis long GRBs and off-axis TDEs at gigahertz frequency \citep[see table~1 and figure~3 in][]{Metzger2015}.

If indeed VAST J175036.1$-$181454 is an off-axis long GRB, the average inverse beaming fraction can then be estimated using the method introduced in \citet{Levinson2002}\footnote{$\tau_i$, the time at which the afterglow becomes isotropic, has a weak dependence with the inverse beaming fraction $f^{-1}_{\text{b}} \propto \tau_i^{7/20}$ so we do not consider this parameter in our estimation (giving it the default assumed value of 3\,yr)}:
\begin{equation}\label{eq:inversebeamingfraction}
\begin{split}
\langle f^{-1}_{\text{b}} \rangle &\approx  70\,N\,\bigg(\frac{\mathcal{R}_\textrm{on-axis}}{0.3\,\textrm{Gpc}^{-3}\,\textrm{yr}^{-1}}\bigg)^{-1}\,\bigg(\frac{E_{\theta,\textrm{kin}}}{10^{51}\,\textrm{erg}}\bigg)^{-11/6}
\\ & \phantom{{}={}} \times \, \bigg(\frac{n}{10\,\textrm{cm}^{-3}}\bigg)^{-19/24}\,\bigg(\frac{\epsilon_\textrm{e}}{0.1}\bigg)^{-3/2}\,\bigg(\frac{\epsilon_\textrm{B}}{0.001}\bigg)^{-9/8},
\end{split}
\end{equation}
where $N$ is the number of orphan afterglows found in the search above the minimum flux-density threshold of 2\,mJy at gigahertz frequency (this is the all-sky number so we had corrected our observed count by the sky fraction $\Delta\Omega$ and survey completeness $C_\textrm{2mJy}$), $\mathcal{R}_\textrm{on-axis}$ is the observed rate for on-axis long GRBs in the local Universe, $E_{\theta,\textrm{kin}}$ is the total beam-corrected kinetic energy in the blast wave, $n$ is density of the circumburst medium, $\epsilon_\textrm{e}$ and $\epsilon_\textrm{B}$ are the fractional shock energies in the accelerated electrons and magnetic fields, respectively.
Taking $\mathcal{R}_\textrm{on-axis} = 0.3\,$Gpc$^{-3}$\,yr$^{-1}$ \citep[e.g.,][]{Guetta2005}, $E_{\theta,\textrm{kin}} = 10^{51}\,$erg \citep[e.g.,][]{Frail2001}, $n=10\,$cm$^{-3}$ \citep[e.g.,][]{Ghisellini2009}, $\epsilon_\textrm{e} = 0.1$ \citep[e.g.,][]{Beniamini2017} and $\epsilon_\textrm{B} = 0.001$ \citep[e.g.,][]{Santana2014}, we obtain the average inverse beaming factor $\langle f^{-1}_{\text{b}} \rangle = 860^{+1980}_{-710}$, or equivalently, an average jet opening angle of $\langle \theta_{\textrm{j}} \rangle = 3^{+4}_{-1}\,$deg, and an implied true long GRB rate of $\langle f^{-1}_{\text{b}} \rangle \, \mathcal{R}_\textrm{on-axis} = 260^{+590}_{-210}\,$Gpc$^{-3}$yr$^{-1}$ (uncertainties representing the $1\sigma$ confidence intervals).
While these results are broadly consistent with previous studies \citep[e.g.,][]{Frail2001,Levinson2002,Guetta2005,Gal-Yam2006,Goldstein2016,Mooley2022,Ghirlanda2022}, we caution that our estimate is only a general approximation, since the methodology suffers from being fairly model dependent (e.g., the distribution of possible values for both $E_{\theta,\textrm{kin}}$ and $\epsilon_\textrm{B}$ span a few orders of magnitude and have a strong dependence -- approximately quadratic and linear, respectively -- on the average inverse beaming fraction $\langle f^{-1}_{\textrm{b}} \rangle$) and also because our results still have many uncertainties remaining (e.g., confirmation of the transient photometric redshift, modelling of the microphysical parameters, etc.).

\subsection{Radio Transient Search Strategies}
\begin{figure*}
	\includegraphics[width=0.8\linewidth,clip,trim={15mm 20mm 30mm 35mm}]{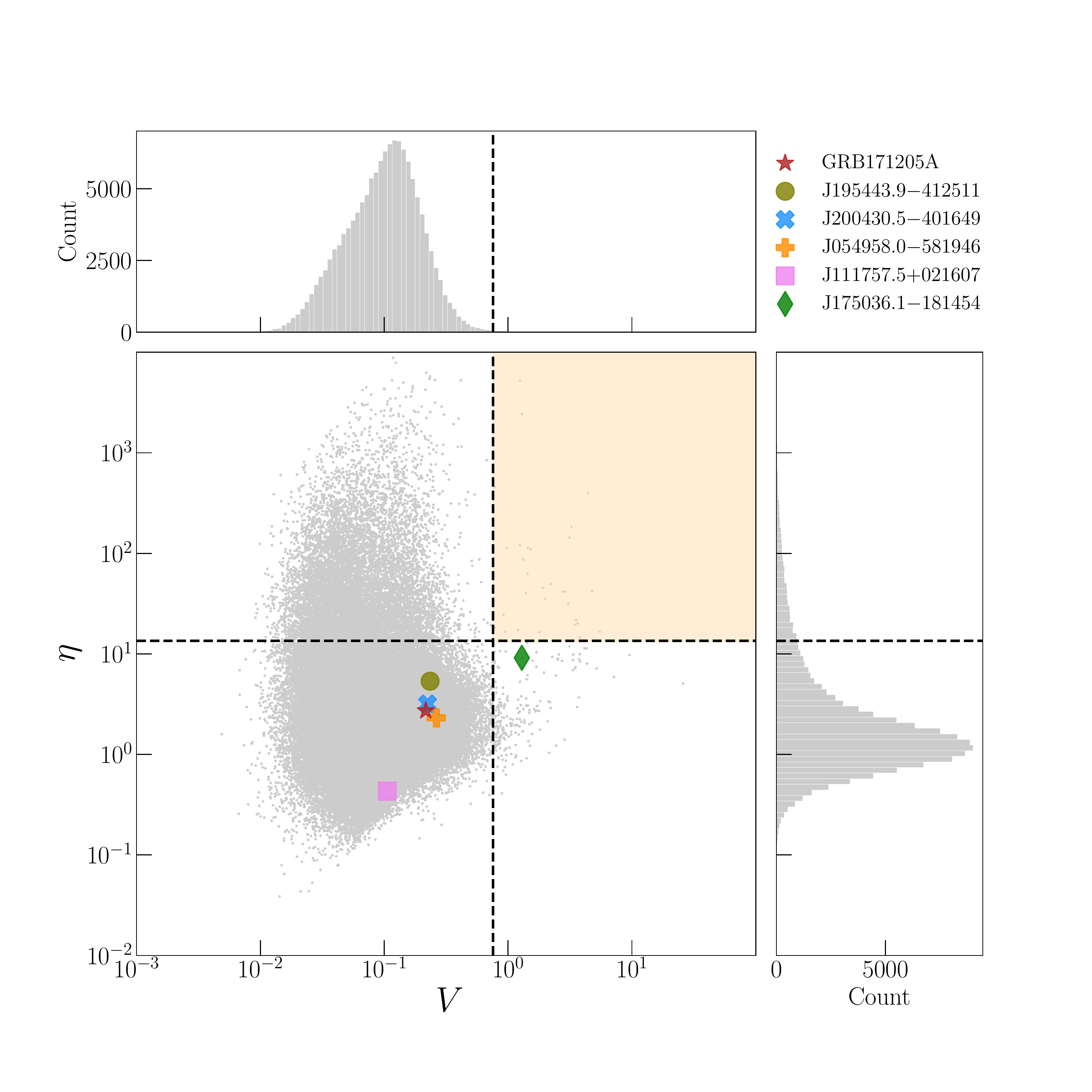}
    \caption{
    Plot of $\eta$ against $V$, two standard variability metrics used to identify variable and transient radio sources in VAST-P.
    The dotted lines are $3\sigma$ thresholds for the variability statistics, above which (represented by the light yellow shaded region) a radio source is identified as a strong variable or transient: $\eta = 13.43$ (horizontal line) and $V = 76$ per cent.
    Each grey point represents a source in VAST-P.
    Candidates from this work and GRB 171205A, which motivated this work, are represented by different colours and markers as shown in the legend on the upper right of the plot.
    }
    \label{fig:eta-v}
\end{figure*}

VAST-P is one of the first radio surveys to have repeated, regular coverage of a significant fraction of the sky, allowing for comprehensive light curves spanning years to be built for many sources.
This has allowed for the sensitivity of different search techniques to variables and transients with different types of light curves -- including those with random/stochastic variations, a single pulse/spike, slow power-law rises/decays -- to be explored in depth.
In particular, we compare the utility of the standard $\eta-V$ statistical approach with our matched-filter approach in variability and transient surveys.
Figure \ref{fig:eta-v} shows a plot of the $\eta$ and $V$ values for all sources in the VAST-P footprint: the shaded area represents the region in variability parameter space exceeding the $3\sigma$ thresholds for both $\eta$ and $V$, while the overlaid markers represent the five orphan afterglow candidates presented in \textsection\ref{sec:results} (as well as GRB 171205A).

Sources with light curves that exhibit random, stochastic variations are now routinely recovered with the $\eta-V$ approach, provided these variations exceed the statistical thresholds defined by the search parameters.
These light curves make up the majority of the 71 sources in the shaded region of Figure \ref{fig:eta-v} and also all the variables identified in the search performed on VAST-P1 \citep{Murphy2021}.
They are typical of those that may be expected in radio stars, pulsars as well as strong scintillators.
For sources with single pulse or spike light curves, which would be observed from a transient whose timescale for rise and decay are faster than the sampling cadence of the survey, the $\eta-V$ approach would be able to recover a subset of the most luminous sources.
If we assume the same noise characteristics and number of observations as VAST-P (low-band), a light curve with a single spike would only be detected by an $\eta-V$ search with $3\sigma$ thresholds on $\eta$ and $V$ if the spike was above ${\sim} 3\,$mJy (i.e., a $12\sigma$ detection).
For a transient with peak spectral luminosity of ${\sim} 10^{30}$\,erg\,s$^{-1}$\,Hz$^{-1}$ at gigahertz frequency (typical of a bright, long GRB radio afterglow), this equates to a detectability radius out to redshift $z \sim 0.1$ or luminosity distance $D_{\text{L}} \sim 500$\,Mpc.
This suggests only rare transients, which are both bright and nearby, would be detectable in a search with similar survey and variability parameters, possibly leading to a biased sample of extragalactic transients.

The matched-filter approach for finding variables and transients is limited by the functional forms or templates that the light curves need to fit.
This makes finding both sources with random, stochastic variations and those with a single spike in their light curve very difficult for a PL/SBPL matched-filter.
However, if an appropriate functional template is chosen, they allow the identification of variables and transients with much lower levels of significance (based on traditional variability metrics) and closer to the noise.
For example, unlike a requirement for a $12\sigma$ detection, a matched-filter approach using a top-hat template had been introduced to find single-spike transients close to the noise \citep{Feng2017}; although the study was implemented at much lower frequencies ($<200\,$MHz) and did not find any transients.
In this work, the PL/SBPL matched-filter was designed to find synchrotron transients, in particular, orphan GRB afterglows, with temporal rise and decays following power laws.
This was motivated by the slow temporal evolution in GRB 171205A at low frequencies at late times \citep{Leung2021}.
Just like GRB 171205A, all five orphan afterglow candidates identified with the PL/SBPL matched-filter in this work would not have been detected in an $\eta-V$ search using $3\sigma$ thresholds\footnote{We note that with more relaxed $2\sigma$ thresholds, which would yield 304 variable and transient candidates, VAST J175036.1$-$181454 would have been detected.} (see Figure~\ref{fig:eta-v}).

\begin{figure}
    \includegraphics[width=\linewidth,clip,trim={0mm 0mm 0mm 0mm}]{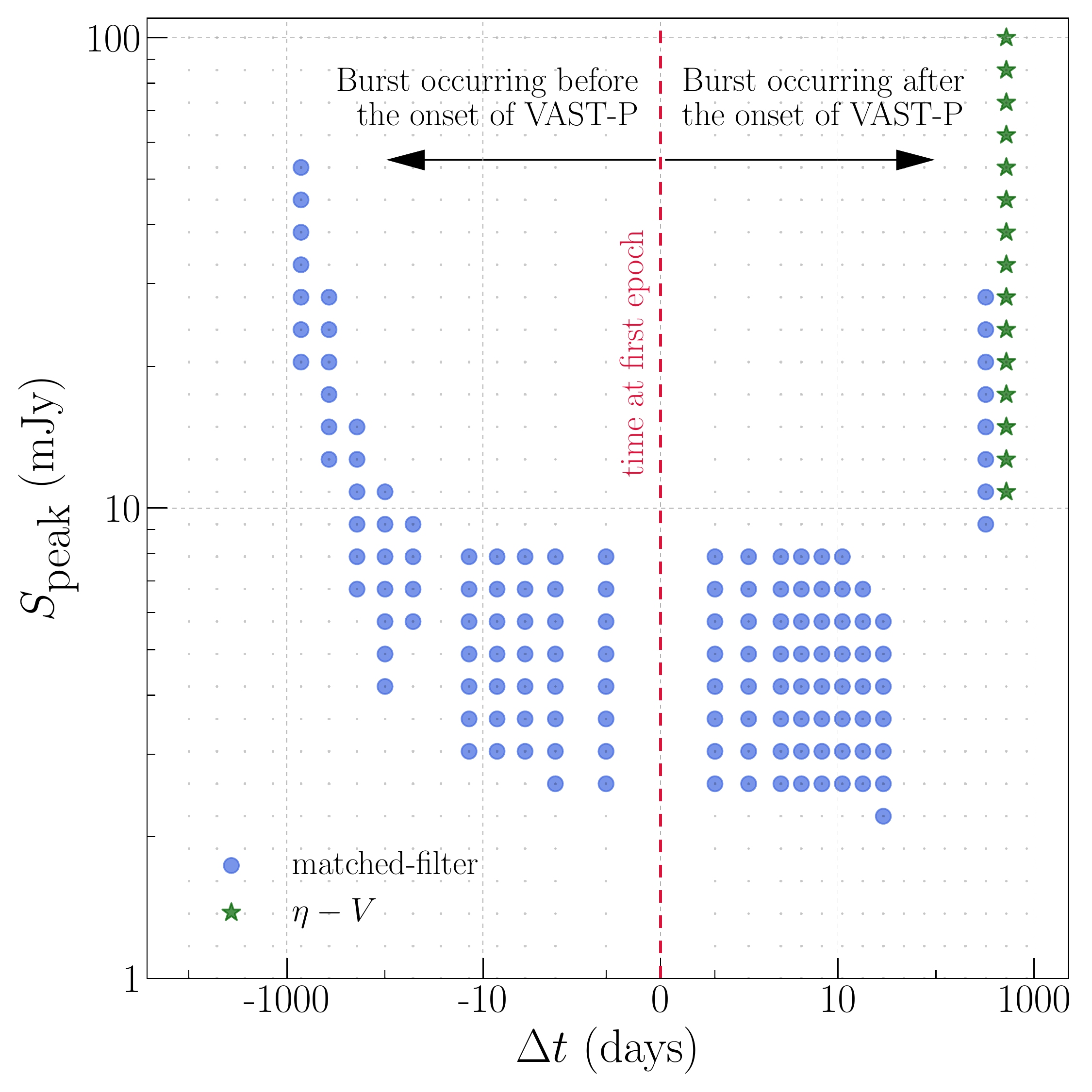}
    \caption{
    A grid of 900 light curves, generated using a SBPL functional template, varying logarithmically in its input parameters $S_\textrm{peak}$ (the maximum flux density) and $\Delta t$ (time elapsed between the first measurement and the time of a burst event).
    The blue circular markers represent the light curves that would be recovered through a PL/SBPL matched-filter search, while the green star markers represent the light curves that would be recovered through a search using the $\eta, V$ variability metrics at $3\sigma$ thresholds.
    }
    \label{fig:etav_mf}
\end{figure}

As an illustrative example, we compare the transient yields of the matched-filter approach to the standard $\eta-V$ approach for sources with SBPL light curves in Figure~\ref{fig:etav_mf}.
To do this, we generated 900 light curves, varying the $S_\textrm{peak}$ (flux density at the light-curve peak) and $\Delta t$ (time elapsed between the first measurement and the time of a burst event) parameters over a logarithmic grid; the $S_\textrm{peak}$ parameter varied from 1 to 100\,mJy and the $\Delta t$ parameter from $-10\,000$ to $1\,000$\,days.
We fixed the time of peak to $t_\textrm{peak} = 100\,$days \citep[consistent with predictions from afterglow models with typical microphysical parameters;][]{Sari1998}, the rise slope to $\delta_1 = 1$ (the expected rise slope for an on-axis GRB in a wind environment; while this parameter varies considerably depending on the inclination angle and jet model for an off-axis GRB, we found the effect on our simulation results below is negligible) and the decay slope to $\delta_2 = -1.5$ (corresponding to $p=2.3$ for a GRB exploding in a wind environment).
For each light curve, we replicated the same observing cadence as our VAST-P observations (low-band) (see Table~\ref{tab:vast_epochs}) and added two sources of noise (see \textsection\ref{sec:observations}): (a) from the scatter in the VAST-P flux-density scale, and (b) from the image noise.

Figure~\ref{fig:etav_mf} shows the light curves that would be recovered through a PL/SBPL matched-filter search (blue circular markers) compared to those that would be recovered through a search using the $\eta, V$ variability metrics at $3\sigma$ thresholds (green star markers).
The results show that the $\eta - V$ search performs well in finding new and bright afterglow events, but is otherwise hampered by the mean flux-density normalisation factor in the modulation index (Equation \ref{eq:var1}) for events with many more data points.
Alternatively, the PL/SBPL matched-filter is able to recover sources occurring much earlier and with lower brightness.
Its performance, however, is hampered for sources with peak brightness greater than 10\,mJy due to the large scatter in the flux-density scale (but this is expected to improve in future ASKAP surveys).
While the results suggest the matched-filter approach would be unable to recover bursts occurring greater than 1\,000\,days before the first epoch, we did not account for possible flattening in the light-curve decay, which may be expected as the afterglow enters the deep-Newtonian regime \citep{Sironi2013}.
Our illustrative example here therefore suggests the $\eta-V$ approach would be sensitive to more bursty light curves (such as radio emission from a new burst), while the matched-filter approach would be more sensitive to lower signal and slowly evolving light curves (such as radio emission from a burst occurring many years prior), which is consistent with our expectations.

Although the matched-filter approach is effective in finding transient candidates exhibiting PL/SBPL light-curve structure, confirmation and characterisation of these transient candidates rely on complementary information.
In \textsection\ref{sec:search}, the 193 candidates from the PL/SBPL matched-filter cuts were narrowed down to five only after the use of multi-frequency radio data (ruling out ${\sim} 30$ per cent of candidates) and archival radio data (ruling out ${\sim} 40$ per cent of candidates), among other auxiliary information including, but not limited to, multi-wavelength and/or redshift information.
This work therefore highlights the need for (a) multiple, complementary approaches to variability searches in addition to the use of standard variability metrics (in particular, we showed the addition of the matched-filter approach would benefit afterglow and synchrotron transient recovery), and (b) the availability of both archival and contemporaneous, complementary, multi-frequency/wavelength data, in order to maximise the scientific value of future variability and transient surveys.

\section*{Acknowledgements}
We thank the anonymous referee for their useful comments that improved the quality of this manuscript. 
We also thank Jess Broderick, Dougal Dobie, Marcello Giroletti, Marcus Lower, Ilya Mandel, Hao Qiu and Om Sharan Salafia for useful discussions. 
JL is supported by Australian Government Research Training Program Scholarships.
TM acknowledges the support of the Australian Research Council through grant DP190100561.
GG acknowledges the support of the PRIN-INAF “Towards the SKA and CTA era: discovery,
localisation, and physics of transient sources” grant (1.05.01.88.06) and PRIN-MUR 2017 grant (20179ZF5KS).
DK is supported by National Science Foundation grant AST-1816492. 
Parts of this research were conducted by the Australian Research Council Centre of Excellence for Gravitational Wave Discovery (OzGrav), through project number CE170100004.

This scientific work uses data obtained from Inyarrimanha Ilgari Bundara / the Murchison Radio-astronomy Observatory. We acknowledge the Wajarri Yamaji People as the Traditional Owners and native title holders of the Observatory site. CSIRO's ASKAP radio telescope is part of the Australia Telescope National Facility (\url{https://ror.org/05qajvd42}). Operation of ASKAP is funded by the Australian Government with support from the National Collaborative Research Infrastructure Strategy. ASKAP uses the resources of the Pawsey Supercomputing Research Centre. Establishment of ASKAP, Inyarrimanha Ilgari Bundara, the CSIRO Murchison Radio-astronomy Observatory and the Pawsey Supercomputing Research Centre are initiatives of the Australian Government, with support from the Government of Western Australia and the Science and Industry Endowment Fund.
The Australia Telescope Compact Array is part of the Australia Telescope National Facility (\url{https://ror.org/05qajvd42}) which is funded by the Australian Government for operation as a National Facility managed by CSIRO.
We acknowledge the Gomeroi people as the Traditional Owners of the Observatory site.

This research was supported by the Sydney Informatics Hub (SIH), a core research facility at the University of Sydney.
This work was also supported by software support resources awarded under the Astronomy Data and Computing Services (ADACS) Merit Allocation Program. ADACS is funded from the Astronomy National Collaborative Research Infrastructure Strategy (NCRIS) allocation provided by the Australian Government and managed by Astronomy Australia Limited (AAL).

This publication makes use of data products from the \textit{Wide-field Infrared Survey Explorer}, which is a joint project of the University of California, Los Angeles, and the Jet Propulsion Laboratory/California Institute of Technology, funded by the National Aeronautics and Space Administration.

The Legacy Surveys consist of three individual and complementary projects: the Dark Energy Camera Legacy Survey (DECaLS; Proposal ID \#2014B-0404; PIs: David Schlegel and Arjun Dey), the Beijing-Arizona Sky Survey (BASS; NOAO Prop. ID \#2015A-0801; PIs: Zhou Xu and Xiaohui Fan), and the Mayall z-band Legacy Survey (MzLS; Prop. ID \#2016A-0453; PI: Arjun Dey). DECaLS, BASS and MzLS together include data obtained, respectively, at the Blanco telescope, Cerro Tololo Inter-American Observatory, NSF’s NOIRLab; the Bok telescope, Steward Observatory, University of Arizona; and the Mayall telescope, Kitt Peak National Observatory, NOIRLab. Pipeline processing and analyses of the data were supported by NOIRLab and the Lawrence Berkeley National Laboratory (LBNL). The Legacy Surveys project is honored to be permitted to conduct astronomical research on Iolkam Du’ag (Kitt Peak), a mountain with particular significance to the Tohono O’odham Nation.
NOIRLab is operated by the Association of Universities for Research in Astronomy (AURA) under a cooperative agreement with the National Science Foundation. LBNL is managed by the Regents of the University of California under contract to the U.S. Department of Energy.
This project used data obtained with the Dark Energy Camera (DECam), which was constructed by the Dark Energy Survey (DES) collaboration. Funding for the DES Projects has been provided by the U.S. Department of Energy, the U.S. National Science Foundation, the Ministry of Science and Education of Spain, the Science and Technology Facilities Council of the United Kingdom, the Higher Education Funding Council for England, the National Center for Supercomputing Applications at the University of Illinois at Urbana-Champaign, the Kavli Institute of Cosmological Physics at the University of Chicago, Center for Cosmology and Astro-Particle Physics at the Ohio State University, the Mitchell Institute for Fundamental Physics and Astronomy at Texas A\&M University, Financiadora de Estudos e Projetos, Fundacao Carlos Chagas Filho de Amparo, Financiadora de Estudos e Projetos, Fundacao Carlos Chagas Filho de Amparo a Pesquisa do Estado do Rio de Janeiro, Conselho Nacional de Desenvolvimento Cientifico e Tecnologico and the Ministerio da Ciencia, Tecnologia e Inovacao, the Deutsche Forschungsgemeinschaft and the Collaborating Institutions in the Dark Energy Survey. The Collaborating Institutions are Argonne National Laboratory, the University of California at Santa Cruz, the University of Cambridge, Centro de Investigaciones Energeticas, Medioambientales y Tecnologicas-Madrid, the University of Chicago, University College London, the DES-Brazil Consortium, the University of Edinburgh, the Eidgenossische Technische Hochschule (ETH) Zurich, Fermi National Accelerator Laboratory, the University of Illinois at Urbana-Champaign, the Institut de Ciencies de l’Espai (IEEC/CSIC), the Institut de Fisica d’Altes Energies, Lawrence Berkeley National Laboratory, the Ludwig Maximilians Universitat Munchen and the associated Excellence Cluster Universe, the University of Michigan, NSF’s NOIRLab, the University of Nottingham, the Ohio State University, the University of Pennsylvania, the University of Portsmouth, SLAC National Accelerator Laboratory, Stanford University, the University of Sussex, and Texas A\&M University.
BASS is a key project of the Telescope Access Program (TAP), which has been funded by the National Astronomical Observatories of China, the Chinese Academy of Sciences (the Strategic Priority Research Program “The Emergence of Cosmological Structures” Grant \# XDB09000000), and the Special Fund for Astronomy from the Ministry of Finance. The BASS is also supported by the External Cooperation Program of Chinese Academy of Sciences (Grant \# 114A11KYSB20160057), and Chinese National Natural Science Foundation (Grant \# 12120101003, \# 11433005).
The Legacy Survey team makes use of data products from the Near-Earth Object Wide-field Infrared Survey Explorer (NEOWISE), which is a project of the Jet Propulsion Laboratory/California Institute of Technology. NEOWISE is funded by the National Aeronautics and Space Administration.
The Legacy Surveys imaging of the DESI footprint is supported by the Director, Office of Science, Office of High Energy Physics of the U.S. Department of Energy under Contract No. DE-AC02-05CH1123, by the National Energy Research Scientific Computing Center, a DOE Office of Science User Facility under the same contract; and by the U.S. National Science Foundation, Division of Astronomical Sciences under Contract No. AST-0950945 to NOAO.
The Photometric Redshifts for the Legacy Surveys (PRLS) catalogue used in this paper was produced thanks to funding from the U.S. Department of Energy Office of Science, Office of High Energy Physics via grant DE-SC0007914.

The Pan-STARRS1 Surveys (PS1) and the PS1 public science archive have been made possible through contributions by the Institute for Astronomy, the University of Hawaii, the Pan-STARRS Project Office, the Max-Planck Society and its participating institutes, the Max Planck Institute for Astronomy, Heidelberg and the Max Planck Institute for Extraterrestrial Physics, Garching, The Johns Hopkins University, Durham University, the University of Edinburgh, the Queen's University Belfast, the Harvard-Smithsonian Center for Astrophysics, the Las Cumbres Observatory Global Telescope Network Incorporated, the National Central University of Taiwan, the Space Telescope Science Institute, the National Aeronautics and Space Administration under Grant No. NNX08AR22G issued through the Planetary Science Division of the NASA Science Mission Directorate, the National Science Foundation Grant No. AST-1238877, the University of Maryland, Eotvos Lorand University (ELTE), the Los Alamos National Laboratory, and the Gordon and Betty Moore Foundation.

Based on observations made with ESO Telescopes at the La Silla or Paranal Observatories under programme ID(s) 179.B-2002(B), 179.B-2002(A), 179.B-2002(C).

This research has made use of the CIRADA cutout service at URL \url{cutouts.cirada.ca}, operated by the Canadian Initiative for Radio Astronomy Data Analysis (CIRADA). CIRADA is funded by a grant from the Canada Foundation for Innovation 2017 Innovation Fund (Project 35999), as well as by the Provinces of Ontario, British Columbia, Alberta, Manitoba and Quebec, in collaboration with the National Research Council of Canada, the US National Radio Astronomy Observatory and Australia’s Commonwealth Scientific and Industrial Research Organisation. 

\section*{Data Availability}
The ASKAP data used in this paper can be accessed through the CSIRO ASKAP Science Data Archive (CASDA)\footnote{\url{https://data.csiro.au/dap/public/casda/casdaSearch.zul}}. Specifically the data from the RACS and VAST projects are available under the codes AS110 and AS107, respectively.
The ATCA data used in this paper can be accessed through the Australia Telescope Online Archive (ATOA)\footnote{\url{https://atoa.atnf.csiro.au/query.jsp}} under the project code C3363.
Reasonable requests for other auxiliary datasets can be accommodated for via email to the corresponding author.

\bibliographystyle{mnras}
\bibliography{bibliography}
\newpage
\appendix
\section{Measurement Tables}
\label{appendix:measurements}
This appendix section provides the full table of radio measurements for all five orphan afterglow candidates investigated in \textsection\ref{sec:results}, including the measurements from the follow-up ATCA observations carried out in this work.

\begin{table}
\caption{
Radio observations of the three orphan afterglow candidates we obtained from performing a SBPL fit to the source light curves.
Columns 1 through 4 show the date of observation, the radio survey or observing telescope, the central frequency of the radio image, and the flux-density measurements for the observation.
Epochs of ATCA follow-up observations detailed in \textsection\ref{ssec:follow-up} separated by dashed horizontal lines.
For the ATCA 2022 September 8 observation, images produced from using the full $2\,048$\, MHz bandwidth is indicated with the dagger $(^{\dagger})$ symbol; all other images from that epoch are produced from using a sub-band with $512$\,MHz bandwidth.
For a non-detection, a $5\sigma$ limit is reported and, where applicable, a measurement from forced extraction (see \textsection\ref{ssec:pipeline}) is given in parenthesis.
The reported uncertainties are purely statistical and these were the uncertainties used for calculations in our pipelines.
The systematic errors (not factored into our quoted numbers) are $\lesssim 5$ per cent for ATCA \citep[e.g.][]{Reynolds1994,Tingay2003}, ${\sim} 10$ per cent for VLASS \citep{Lacy2022}, typically ${\sim} 7$ per cent but \textit{up to} ${\sim} 30$ per cent for ASKAP \citep{McConnell2020,Duchesne2023}.
}
\label{tab:sbpl_measurements1}
\begin{tabular}{lccc}
\hline\hline
Date (UT) & Survey or Telescope & $\nu$ (GHz) & $S_\nu$ (mJy) \\
\hline \multicolumn{4}{l}{\textbf{SBPL Candidate 1: VAST J195443.9$-$412511}} \\ \hline
2019 Apr 30 & VAST-P & 0.887 & $< 3.64\ (2.42 \pm 0.73)$\\ 
2019 Aug 28 & \textquotedbl\textquotedbl & \textquotedbl\textquotedbl & $5.11 \pm 0.37$\\ 
2019 Oct 30 & \textquotedbl\textquotedbl & \textquotedbl\textquotedbl & $6.26 \pm 0.39$\\ 
2020 Jan 26 & \textquotedbl\textquotedbl & \textquotedbl\textquotedbl & $5.68 \pm 0.37$\\ 
2020 Jan 27 & \textquotedbl\textquotedbl & \textquotedbl\textquotedbl & $5.55 \pm 0.35$\\ 
2020 Jun 21 & \textquotedbl\textquotedbl & \textquotedbl\textquotedbl & $4.66 \pm 0.43$\\ 
2020 Jul 4 & FLASH-P & 0.855 & $6.39 \pm 0.02$\\ 
2020 Aug 30 & VAST-P & 0.887 & $5.32 \pm 0.49$\\ 
2021 Jul 22 & \textquotedbl\textquotedbl & \textquotedbl\textquotedbl & $4.14 \pm 0.32$\\ 
2021 Jul 30 & \textquotedbl\textquotedbl & 1.367 & $4.44 \pm 0.31$\\ 
2021 Aug 22 & \textquotedbl\textquotedbl & 0.887 & $4.28 \pm 0.31$\\ 
2021 Sep 24 & \textquotedbl\textquotedbl & 1.367 & $4.45 \pm 0.26$\\ 
2021 Nov 20 & \textquotedbl\textquotedbl & \textquotedbl\textquotedbl & $4.36 \pm 0.26$ \vspace{1mm} \\
\hdashline \vspace{-2.5mm} \\
2022 May 16 & ATCA & 2.100 & $3.99 \pm 0.23$\\ 
\textquotedbl\textquotedbl & \textquotedbl\textquotedbl & 5.500 & $3.71 \pm 0.09$\\ 
\textquotedbl\textquotedbl & \textquotedbl\textquotedbl & 9.000 & $3.44 \pm 0.10$ \vspace{1mm} \\
\hdashline \vspace{-2.5mm} \\
2022 Sep 8 & ATCA & 1.332 & $3.87 \pm 0.33$\\ 
\textquotedbl\textquotedbl & \textquotedbl\textquotedbl & 1.844 & $3.95 \pm 0.18$\\ 
\textquotedbl\textquotedbl & \textquotedbl\textquotedbl & \phantom{$^0$}2.100$^{\dagger}$ & $3.76 \pm 0.13$\\ 
\textquotedbl\textquotedbl & \textquotedbl\textquotedbl & 2.356 & $3.57 \pm 0.16$\\ 
\textquotedbl\textquotedbl & \textquotedbl\textquotedbl & 2.868 & $3.18 \pm 0.20$\\ 
\textquotedbl\textquotedbl & \textquotedbl\textquotedbl & 4.732 & $2.72 \pm 0.19$\\ 
\textquotedbl\textquotedbl & \textquotedbl\textquotedbl & 5.244 & $2.48 \pm 0.16$\\ 
\textquotedbl\textquotedbl & \textquotedbl\textquotedbl & \phantom{$^0$}5.500$^{\dagger}$ & $2.62 \pm 0.11$\\ 
\textquotedbl\textquotedbl & \textquotedbl\textquotedbl & 5.756 & $2.43 \pm 0.17$\\ 
\textquotedbl\textquotedbl & \textquotedbl\textquotedbl & 6.268 & $2.29 \pm 0.20$\\ 
\textquotedbl\textquotedbl & \textquotedbl\textquotedbl & 8.232 & $1.86 \pm 0.14$\\ 
\textquotedbl\textquotedbl & \textquotedbl\textquotedbl & 8.744 & $1.59 \pm 0.15$\\ 
\textquotedbl\textquotedbl & \textquotedbl\textquotedbl & \phantom{$^0$}9.000$^{\dagger}$ & $1.65 \pm 0.08$\\ 
\textquotedbl\textquotedbl & \textquotedbl\textquotedbl & 9.256 & $1.57 \pm 0.17$\\ 
\textquotedbl\textquotedbl & \textquotedbl\textquotedbl & 9.768 & $1.38 \pm 0.16$\\ 
\hline \multicolumn{4}{r}{\textit{continued on the next page}}
\end{tabular}
\end{table}
\begin{table}
\caption*{\textit{continued from the previous page}}
\label{tab:sbpl_measurements2}
\begin{tabular}{lccc}
\hline\hline
Date (UT) & Survey or Telescope & $\nu$ (GHz) & $S_\nu$ (mJy) \\

\hline \multicolumn{4}{l}{\textbf{SBPL Candidate 2: VAST J200430.5$-$401649}} \\ \hline
2018 Feb 3 & VLASS & 3.000 & $< 0.66$\\ 
2019 Apr 30 & VAST-P & 0.887 & $< 2.78\ (2.55 \pm 0.56)$\\ 
2019 Aug 28 & VAST-P & 0.887 & $4.45 \pm 0.46$\\ 
2019 Oct 30 & \textquotedbl\textquotedbl & \textquotedbl\textquotedbl & $5.16 \pm 0.50$\\ 
2020 Jan 26 & \textquotedbl\textquotedbl & \textquotedbl\textquotedbl & $4.68 \pm 0.51$\\ 
2020 Jan 27 & \textquotedbl\textquotedbl & \textquotedbl\textquotedbl & $4.70 \pm 0.52$\\ 
2020 Jun 21 & \textquotedbl\textquotedbl & \textquotedbl\textquotedbl & $3.66 \pm 0.44$\\ 
2020 Jul 4 & FLASH-P & 0.855 & $4.31 \pm 0.03$\\ 
2020 Aug 30 & VAST-P & 0.887 & $4.10 \pm 0.53$\\ 
2020 Nov 9 & VLASS & 3.000 & $< 0.82$\\ 
2021 Jul 22 & VAST-P & 0.887 & $3.40 \pm 0.45$\\ 
2021 Jul 30 & \textquotedbl\textquotedbl & 1.367 & $2.58 \pm 0.31$\\ 
2021 Aug 22 & \textquotedbl\textquotedbl & 0.887 & $2.91 \pm 0.43$\\ 
2021 Sep 24 & \textquotedbl\textquotedbl & 1.367 & $2.47 \pm 0.34$\\ 
2021 Nov 20 & \textquotedbl\textquotedbl & \textquotedbl\textquotedbl & $2.40 \pm 0.30$ \vspace{1mm} \\
\hdashline \vspace{-2.5mm} \\
2022 May 16 & ATCA & 2.100 & $2.19 \pm 0.54$\\ 
\textquotedbl\textquotedbl & \textquotedbl\textquotedbl & 5.500 & $0.92 \pm 0.08$\\ 
\textquotedbl\textquotedbl & \textquotedbl\textquotedbl & 9.000 & $0.54 \pm 0.08$\\ 

\hline \multicolumn{4}{l}{\textbf{SBPL Candidate 3: VAST J054958.0$-$581946}} \\ \hline
2019 May 4 & VAST-P & 0.887 & $< 1.20\ (0.73 \pm 0.24)$\\ 
2019 Aug 27 & \textquotedbl\textquotedbl & \textquotedbl\textquotedbl & $1.96 \pm 0.34$\\ 
2019 Oct 29 & \textquotedbl\textquotedbl & \textquotedbl\textquotedbl & $1.82 \pm 0.30$\\ 
2019 Oct 31 & \textquotedbl\textquotedbl & \textquotedbl\textquotedbl & $1.62 \pm 0.29$\\ 
2019 Dec 19 & \textquotedbl\textquotedbl & \textquotedbl\textquotedbl & $< 1.29\ (1.47 \pm 0.26)$\\ 
2020 Jan 10 & \textquotedbl\textquotedbl & \textquotedbl\textquotedbl & $1.30 \pm 0.25$\\ 
2020 Jan 16 & \textquotedbl\textquotedbl & \textquotedbl\textquotedbl & $1.46 \pm 0.26$\\ 
2020 Jan 17 & \textquotedbl\textquotedbl & \textquotedbl\textquotedbl & $1.68 \pm 0.21$\\ 
2020 Jan 18 & \textquotedbl\textquotedbl & \textquotedbl\textquotedbl & $1.75 \pm 0.21$\\ 
2020 Jun 20 & \textquotedbl\textquotedbl & \textquotedbl\textquotedbl & $1.51 \pm 0.21$\\ 
2020 Aug 28 & \textquotedbl\textquotedbl & \textquotedbl\textquotedbl & $< 1.30\ (1.01 \pm 0.26)$\\ 
2021 Jul 24 & \textquotedbl\textquotedbl & \textquotedbl\textquotedbl & $< 0.99\ (0.94 \pm 0.20)$\\ 
2021 Aug 22 & \textquotedbl\textquotedbl & \textquotedbl\textquotedbl & $1.13 \pm 0.19$ \vspace{1mm} \\
\hdashline \vspace{-2.5mm} \\
2022 May 16 & ATCA & 2.100 & $1.07 \pm 0.14$\\ 
\textquotedbl\textquotedbl & \textquotedbl\textquotedbl & 5.500 & $0.65 \pm 0.09$\\ 
\textquotedbl\textquotedbl & \textquotedbl\textquotedbl & 9.000 & $0.51 \pm 0.04$\\ 
\hline
\end{tabular}
\end{table}
\begin{table}
\caption{
Radio observations of the two orphan afterglow candidates we obtained from performing a PL fit to the source light curves.
Columns 1 through 4 show the date of observation, the radio survey or observing telescope, the central frequency of the radio image, and the flux-density measurements for the observation.
Epochs of ATCA follow-up observations detailed in \textsection\ref{ssec:follow-up} separated by dashed horizontal lines.
For the ATCA 2022 March 5 observation, images produced from using the full $2\,048$\, MHz bandwidth is indicated with the dagger $(^{\dagger})$ symbol; all other images from that epoch are produced from using a sub-band with $512$\,MHz bandwidth.
For a non-detection, a $5\sigma$ limit is reported and, where applicable, a measurement from forced extraction (see \textsection\ref{ssec:pipeline}) is given in parenthesis.
The reported uncertainties are purely statistical and these were the uncertainties used for calculations in our pipelines.
The systematic errors (not factored into our quoted numbers) are $\lesssim 5$ per cent for ATCA \citep[e.g.][]{Reynolds1994,Tingay2003}, ${\sim} 10$ per cent for VLASS \citep{Lacy2022}, typically ${\sim} 7$ per cent but \textit{up to} ${\sim} 30$ per cent for ASKAP \citep{McConnell2020,Duchesne2023}.
}
\label{tab:pl_measurements1}
\begin{tabular}{lccc}
\hline 
\textit{continued in the next column} &&&
\end{tabular}
\end{table}
\begin{table}
\caption*{\textit{continued from the previous column}}
\label{tab:pl_measurements2}
\begin{tabular}{lccc}
\hline\hline
Date (UT) & Survey or Telescope & $\nu$ (GHz) & $S_\nu$ (mJy) \\

\hline \multicolumn{4}{l}{\textbf{PL Candidate 1: VAST J111757.5$+$021607}} \\ \hline
1998 Jul & FIRST & 1.400 & $< 0.78$\\ 
2017 Dec 31 & VLASS & 3.000 & $1.19 \pm 0.38$\\ 
2019 Aug 28 & VAST-P & 0.887 & $2.48 \pm 0.27$\\ 
2019 Oct 30 & \textquotedbl\textquotedbl & \textquotedbl\textquotedbl & $2.30 \pm 0.29$\\ 
2019 Dec 19 & \textquotedbl\textquotedbl & \textquotedbl\textquotedbl & $2.36 \pm 0.30$\\ 
2020 Jan 10 & \textquotedbl\textquotedbl & \textquotedbl\textquotedbl & $< 4.15\ (1.67 \pm 0.83)$\\ 
2020 Jan 16 & \textquotedbl\textquotedbl & \textquotedbl\textquotedbl & $2.42 \pm 0.31$\\ 
2020 Jan 24 & \textquotedbl\textquotedbl & \textquotedbl\textquotedbl & $2.27 \pm 0.26$\\ 
2020 Jan 25 & \textquotedbl\textquotedbl & \textquotedbl\textquotedbl & $2.36 \pm 0.27$\\ 
2020 May 2 & \textquotedbl\textquotedbl & \textquotedbl\textquotedbl & $2.20 \pm 0.35$\\ 
2020 Jun 20 & \textquotedbl\textquotedbl & \textquotedbl\textquotedbl & $2.12 \pm 0.25$\\ 
2020 Aug 11 & VLASS & 3.000 & $< 0.88$\\ 
2020 Aug 29 & VAST-P & 0.887 & $2.30 \pm 0.32$\\ 
2021 Jul 23 & \textquotedbl\textquotedbl & \textquotedbl\textquotedbl & $1.93 \pm 0.25$\\ 
2021 Aug 22 & \textquotedbl\textquotedbl & \textquotedbl\textquotedbl & $2.00 \pm 0.24$\\ 

\hline \multicolumn{4}{l}{\textbf{PL Candidate 2: VAST J175036.1$-$181454}} \\ \hline
2019 Jun 30 & VLASS & 3.000 & $< 0.84$\\ 
2019 Aug 28 & VAST-P & 0.887 & $< 1.26\ (0.27 \pm 0.25)$\\ 
2019 Oct 29 & \textquotedbl\textquotedbl & \textquotedbl\textquotedbl & $< 1.35\ (-0.01 \pm 0.27)$\\ 
2019 Oct 30 & \textquotedbl\textquotedbl & \textquotedbl\textquotedbl & $< 1.21\ (-0.23 \pm 0.24)$\\ 
2020 Jan 11 & \textquotedbl\textquotedbl & \textquotedbl\textquotedbl & $< 1.15\ (0.57 \pm 0.23)$\\ 
2020 Jan 18 & \textquotedbl\textquotedbl & \textquotedbl\textquotedbl & $< 1.23\ (0.25 \pm 0.25)$\\ 
2020 Jan 19 & \textquotedbl\textquotedbl & \textquotedbl\textquotedbl & $< 1.16\ (0.37 \pm 0.23)$\\ 
2020 May 3 & \textquotedbl\textquotedbl & \textquotedbl\textquotedbl & $< 1.18\ (0.03 \pm 0.24)$\\ 
2020 Jun 20 & \textquotedbl\textquotedbl & \textquotedbl\textquotedbl & $< 1.18\ (0.29 \pm 0.24)$\\ 
2020 Aug 29 & \textquotedbl\textquotedbl & \textquotedbl\textquotedbl & $< 1.51\ (0.55 \pm 0.30)$\\ 
2021 Jul 23 & \textquotedbl\textquotedbl & \textquotedbl\textquotedbl & $1.66 \pm 0.22$\\ 
2021 Jul 29 & \textquotedbl\textquotedbl & 1.367 & $3.06 \pm 0.34$\\ 
2021 Aug 21 & \textquotedbl\textquotedbl & 0.887 & $1.82 \pm 0.20$\\ 
2021 Sep 25 & \textquotedbl\textquotedbl & 1.367 & $3.88 \pm 0.30$\\ 
2021 Nov 20 & \textquotedbl\textquotedbl & \textquotedbl\textquotedbl & $3.13 \pm 0.34$\\ 
2022 Feb 7 & VLASS & 3.000 & $2.39 \pm 0.28$ \vspace{1mm} \\
\hdashline \vspace{-2.5mm} \\
2022 Mar 5 & ATCA & 1.332 & $2.93 \pm 0.23$\\ 
\textquotedbl\textquotedbl & \textquotedbl\textquotedbl & 1.844 & $3.25 \pm 0.13$\\ 
\textquotedbl\textquotedbl & \textquotedbl\textquotedbl & \phantom{$^0$}2.100$^{\dagger}$ & $3.16 \pm 0.09$\\ 
\textquotedbl\textquotedbl & \textquotedbl\textquotedbl & 2.356 & $3.18 \pm 0.11$\\ 
\textquotedbl\textquotedbl & \textquotedbl\textquotedbl & 2.868 & $3.00 \pm 0.11$\\ 
\textquotedbl\textquotedbl & \textquotedbl\textquotedbl & 4.732 & $2.01 \pm 0.17$\\ 
\textquotedbl\textquotedbl & \textquotedbl\textquotedbl & 5.244 & $1.65 \pm 0.15$\\ 
\textquotedbl\textquotedbl & \textquotedbl\textquotedbl & \phantom{$^0$}5.500$^{\dagger}$ & $1.67 \pm 0.08$\\ 
\textquotedbl\textquotedbl & \textquotedbl\textquotedbl & 5.756 & $1.63 \pm 0.16$\\ 
\textquotedbl\textquotedbl & \textquotedbl\textquotedbl & 6.268 & $1.58 \pm 0.17$\\ 
\textquotedbl\textquotedbl & \textquotedbl\textquotedbl & 8.232 & $1.22 \pm 0.19$\\ 
\textquotedbl\textquotedbl & \textquotedbl\textquotedbl & 8.744 & $1.10 \pm 0.16$\\ 
\textquotedbl\textquotedbl & \textquotedbl\textquotedbl & \phantom{$^0$}9.000$^{\dagger}$ & $1.08 \pm 0.08$\\ 
\textquotedbl\textquotedbl & \textquotedbl\textquotedbl & 9.256 & $1.11 \pm 0.16$\\ 
\textquotedbl\textquotedbl & \textquotedbl\textquotedbl & 9.768 & $1.09 \pm 0.15$\\ 
\textquotedbl\textquotedbl & \textquotedbl\textquotedbl & 16.700$^{\dagger}$ & $0.83 \pm 0.11$ \vspace{1mm} \\
\hdashline \vspace{-2.5mm} \\ 
2022 May 16 & ATCA & 2.100 & $2.63 \pm 0.25$\\ 
\textquotedbl\textquotedbl & \textquotedbl\textquotedbl & 5.500 & $1.18 \pm 0.07$\\ 
\textquotedbl\textquotedbl & \textquotedbl\textquotedbl & 9.000 & $0.97 \pm 0.08$\\ 
\hline
\end{tabular}
\end{table}

\bsp	
\label{lastpage}
\end{document}